\documentstyle[epsfig,latexsym,aps,floats,eqsecnum,preprint,
amsfonts]{revtex}

\def\laq{~\raise 0.4ex\hbox{$<$}\kern -0.8em\lower 0.62 
ex\hbox{$\sim$}~}
\def\gaq{~\raise 0.4ex\hbox{$>$}\kern -0.7em\lower 0.62 
ex\hbox{$\sim$}~}

\newcommand{\beq}{\begin{equation}}
\newcommand{\eeq}{\end{equation}}
\newcommand{\bea}{\begin{eqnarray}}
\newcommand{\eea}{\end{eqnarray}}

\def \pa {\partial}

\def \ra {\rightarrow}

\def \b {\beta}
\def \a {\alpha}

\def \Ga {\Gamma}

\def \sg {\sigma}
\def \da {\delta}

\def \r {\rho}
\def \om {\omega}

\begin{document}

\begin{titlepage}

\def\baselinestretch{1.1}

\begin{flushright}
BA-TH/02-451\\
CERN-TH/2002-352 \\
hep-ph/0212112
\end{flushright}

\vspace*{1cm}

\begin{center}
{\large{\bf Constraints on pre-big bang parameter space\\
from CMBR anisotropies}}

\vspace*{1cm}

{\sl V. Bozza$^{a,b,c}$, M. Gasperini$^{d,e}$,
M. Giovannini$^{c}$ and G. Veneziano$^c$}

{\sl $^a$ Dipartimento di Fisica 
``E. R. Caianiello", Universit\`a di
Salerno, 84081 Baronissi, Italy}

{\sl $^b$ INFN, Sezione di
Napoli, Gruppo Collegato di Salerno, Salerno, Italy}

{\sl $^c$  Theoretical Physics Division, CERN, CH-1211
Geneva 23, Switzerland}

{\sl $^d$ Dipartimento di Fisica,
Universit\`a di Bari, Via G. Amendola 173, 70126 Bari, Italy}

{\sl $^e$ INFN, Sezione di Bari, Bari, Italy} 

\begin{abstract}
The so-called curvaton mechanism --a way to convert isocurvature
perturbations into adiabatic ones-- is investigated
 both analytically and numerically in a
 pre-big bang scenario where
 the r\^ole of the 
curvaton is played by a sufficiently massive Kalb--Ramond axion
of superstring theory.
When combined with observations of CMBR anisotropies
 at large and moderate angular scales,
the present analysis allows us to 
constrain quite considerably the parameter space of
the model: in particular, the initial displacement of the axion from the
minimum of its potential and  the rate of evolution of the
compactification volume during pre-big bang inflation. The combination
of theoretical and experimental constraints favours a slightly blue
spectrum of scalar perturbations, and/or a value of the string scale 
in the vicinity of the SUSY-GUT scale.
\end{abstract}

\end{center}
\end{titlepage}

\noindent

\renewcommand{\theequation}{1.\arabic{equation}}
\setcounter{equation}{0}
\section{Introduction} 
\label{Sec1}
At present, the largest-scale temperature fluctuations 
of the Cosmic Microwave Background radiation (CMBR) 
are consistent with a (quasi) scale-invariant spectrum of  Gaussian
primordial curvature fluctuations \cite{cobe,cob2,silk}. 
The analysis 
of the first acoustic oscillations occurring on shorter angular scales 
 adds the information that such 
curvature fluctuations should  be predominantly adiabatic 
\cite{boom,dasi,maxima,archeops}. 
Although sufficiently small amounts of non-Gaussianity and/or 
isocurvature perturbations are not excluded,
the above-mentioned observational features represent 
an important constraint for any scenario trying to model 
the initial stages of our Universe. 
In a previous Letter \cite{us} we  tried to confront the  pre-big
bang  scenario \cite{g1,lwc,g2}
with these constraints. 
The present paper contains a full description of that analysis and
completes it.

Let us recall that, during the pre-big bang phase, 
the quantum fluctuations of all the 
light modes present in the low energy effective action are
parametrically  amplified. Nonetheless, sizeable large-scale
adiabatic fluctuations are not easily produced from the initial
vacuum through the usual mechanism of parametric amplification. 
In particular, both tensor and scalar-metric
fluctuations are amplified with  very steep spectra \cite{GV94,bmggv},  
resulting in  adiabatic modes which are  
far too small to explain the observed level of large-scale
CMBR anisotropies \cite{smoo1}. 

However, not all the primordial spectra of pre-big bang cosmology are 
blue. For instance,
in a pure gravi-dilaton background, the pseudo-scalar
supersymmetric  partner of the
dilaton in the dimensionally-reduced string effective action,
 the so-called Kalb--Ramond axion, emerges from the pre-big bang 
phase with a fluctuation spectrum whose tilt depends on the rate of
change of  the compactification 
volume \cite{cew,clw}. 
Depending on this, the axion tilt can be negative (red spectrum), 
positive (blue spectrum), or zero (scale-invariant spectrum).
However, since the homogeneous (background) component of the 
axion is trivial, such a spectrum does not affect directly the metric, 
hence no curvature perturbation
is generated at this primordial level. In other words, the axion 
perturbations are entropic, isocurvature perturbations. This feature
persists, unfortunately,
 under $S$-duality
transformations \cite{cew,clw}.
If such an axion is massless, or at least light enough not to have 
decayed yet , the induced CMBR fluctuations on large scales 
 can fit the COBE normalization \cite{dgs1,dgs2,dgs3,mas1}, 
but, being neither adiabatic nor  Gaussian
\cite{mvdv}, they are not able to fit the observed  structure of the 
first few  acoustic peaks.
 
A possible way out of this problem \cite{us,lwc,es,lw,Moroi} is  offered
by the alternative scenario of a massive axion, initially displaced
from the minimum of its non-perturbative potential.  In
that case axion perturbations couple to scalar metric perturbations
through the non-vanishing axion's VEV. Eventually, the axion
   relaxes toward the minimum  of the potential
and then, if heavy enough, decays prior to nucleosynthesis.  
During the relaxation
process  the dominant source of energy undergoes a drastic change: 
 it consists of   
the radiation produced at the  end of the pre-big bang
evolution, and later  becomes the pressureless fluid  corresponding
to the damped coherent oscillations of the axion. This non-trivial
evolution
 results in a non-adiabatic
pressure perturbation which, in  turn, is well known \cite{ks,mfb} to
induce  curvature perturbations on constant energy (or comoving)
hypersurfaces even on superhorizon scales.

The interplay of such different sources of inhomogeneity, 
throughout the different stages of the background evolution,
eventually determines the spectral amplitude 
of scalar curvature perturbations right after matter-radiation 
equality, when  all the 
 scales of interest for the CMBR data are still outside 
the horizon. This
 conversion of isocurvature into adiabatic perturbations,
originally suggested in a  different context by Mollerach \cite{mol},
also applies to more general cases \cite{lw,Moroi}.

Depending upon the initial value $\sigma_{\rm i}$ of the  Kalb--Ramond
background,  different 
post-big bang histories are possible. If 
$\bar{\sigma} < \sigma_{\rm i} \ll 1$ in Planck units 
(see below, Eq. (\ref{c1}), for the definition of 
$\bar{\sigma}$), 
the axion  oscillates for a long time  before  becoming 
dominant and eventually decays. For $  \sigma_{\rm i} < \bar{\sigma}$ 
it may never fully dominate
the energy density before decaying. If, instead, $\sg_{\rm i} \gg 1$  the axion
 will dominate before oscillating  and a  slow-roll (low-scale) 
inflationary  phase could take place in that epoch. As we
shall see, of all these possibilities CMBR observations seem
 to  favour the ``natural" one, $\sigma_{\rm i} \sim 1$. In any case, even if
different post-big bang histories will lead 
to different spectral amplitudes of the Bardeen potential,
 adiabatic scalar metric perturbations will
always be present at some level outside the horizon, prior to 
decoupling.

The purpose of the present paper is to report on the calculation of 
 the spectral amplitude of the induced adiabatic metric
perturbations, and on the comparison of the predictions 
of the pre-big bang scenario with the observations coming from the
physics  of the CMBR anisotropies. In order to achieve this goal 
it is mandatory to have a good understanding both of the 
axion relaxation mechanism and of the evolution of the 
inhomogeneities. Hence analytical results will be supported with 
numerical examples and vice versa. We will present, in particular,  a
full derivation of the results for the final adiabatic spectrum of the
Bardeen potential (some of these results have been 
summarized already in \cite{us}). 

The paper is organized as follows. 
In Section II the basic equations describing the 
post-big bang evolution of the inhomogeneities 
and of the background geometry will be introduced.
In Section III the physics of the  different post-big bang histories 
will be analysed. 
In Section IV the evolution of the background and of its perturbations 
will be discussed for the case in which the amplitude of the initial axion
background is 
smaller than $1$ in Planck units, $\sg_{\rm i}<1$. In Section V we will 
discuss the evolution of the system in the complementary 
case $\sg_{\rm i}>1$. Section VI is devoted to the phenomenological 
implications of  the large-scale 
adiabatic perturbations produced through the relaxation 
of the axionic background. The obtained results will be compared with 
observations. Constraints on the pre-big bang parameters 
will be derived. 
Section VII contains our concluding remarks while, in the 
Appendix, a self-contained derivation of the axionic 
spectra produced by the pre-big bang evolution has been included.

\renewcommand{\theequation}{2.\arabic{equation}}
\setcounter{equation}{0}
\section{Background and perturbation equations}

As already mentioned, we shall start our analysis at some 
time $\eta_{\rm i}$ in the post-big bang epoch, assuming that 
the axion field has inherited from the preceding epoch
 appreciable large--scale 
fluctuations, while other sources of energy as well as the metric
are exactly homogeneous. It will also be assumed that, initially,
the dominant source of energy is in the form of radiation.
The post-big bang dynamics takes place, in the present analysis,
 when  the curvature scale has fallen to a sufficiently 
small value (in string units) so that the use of  the low-energy 
effective action
is appropriate. Furthermore, for $\eta >\eta_{\rm i}$ 
 the dilaton is assumed to be frozen already at its present value.
 
Under these assumptions, 
the evolution of the geometry is determined by the Einstein
equations, supplemented by  the conservation 
equations determining the dynamics of  the
sources\footnote{Gravitational units
 $16 \pi G= 1$ 
will be used throughout. When explicitely written in the formulae, 
$M_{\rm P} = (16 \pi G)^{-1/2} = 1.72\times 10^{18}~{\rm GeV}$. In these 
units,
 $\sigma$ is the canonically 
normalized axion field.}:
\begin{eqnarray}
&& R_{\alpha}^{\beta} - \frac{1}{2} \delta_{\alpha}^{\beta} R = 
\frac{1}{2} \biggl( T_{\alpha}^{\beta}(\sigma) 
+ {\cal T}_{\alpha}^{\beta} \biggr),
\label{g1}\\
&& g^{\alpha\beta} \nabla_{\alpha} \nabla_{\beta} \sigma + 
\frac{\partial V}{\partial\sigma} =0,
\label{g2}
\end{eqnarray}
where $T_{\alpha}^{\beta}(\sigma)$ and ${\cal T}_{\alpha}^{\beta}$ are, 
respectively, the energy-momentum tensors of the axionic background 
and of the matter fluid. Notice that the covariant conservation of 
$T_{\alpha}^{\beta}(\sigma)$   is dynamically 
equivalent to  the evolution equation of the axionic field, i.e. 
Eq. (\ref{g2}), and implies, through the contracted Bianchi identities:
\beq
 \nabla_{\alpha} {\cal T}_{\beta}^{\alpha} =0. 
\label{g3}
\eeq
In a conformally flat  background geometry, 
\begin{equation}
ds^2 = a^2(\eta) [ d\eta^2 - d\vec{x}^2], 
\label{lel}
\end{equation}
Eqs. (\ref{g1})--(\ref{g3}) lead to a set of three independent 
equations, whose specific form is dictated by the fluid content 
of the primordial plasma. In the case of 
a radiation fluid we have
\begin{equation}
{\cal T}_{0}^{0} = \rho_{\rm r}, ~~~~~ {\cal T}_{i}^{j} 
= - p_{\rm r} \delta_{i}^{j}, ~~~~~p_{\rm r} 
= \frac{\rho_{\rm r}}{3},
\end{equation}
 and  Eqs. (\ref{g1})--(\ref{g3}) lead to
\begin{eqnarray}
&& {\cal H}' = - \frac{a^2}{6} \biggl[ \rho_{\rm r} 
+ \frac{ {\sigma'}^2}{a^2} - V\biggr],
\label{dyn}\\
&& \sigma'' + 2 {\cal H} \sigma' + a^2 \frac{\partial V}{\partial\sigma} 
=0, \label{kg}\\
&& \rho_{\rm r}' + 4 {\cal H} \rho_{\rm r} =0.
\label{contrad}
\end{eqnarray}
Here the prime denotes the  derivation with 
respect to the conformal time coordinate $\eta$, and 
${\cal H}= (\ln{a})'$. For future convenience we also 
recall that the connection between ${\cal H}$ and the 
Hubble parameter is $ H= {\cal H}/a$. 
The
 effective energy and pressure densities of $\sigma$ will be given by 
\begin{equation}
\rho_{\sigma} = \frac{{\sigma'}^2}{2 a^2} + V, ~~~~~~~~
p_{\sigma} =    \frac{{\sigma'}^2}{2 a^2} - V.
\label{backdef}
\end{equation}
The set of dynamical equations (\ref{dyn})--(\ref{contrad}) is 
supplemented by the Hamiltonian constraint 
\begin{equation}
{\cal H}^2 = \frac{a^2}{6} \biggl[ \rho_{\rm r} + \frac{{\sigma'}^2}{2 a^2} 
+ V\biggr],
\label{ham}
\end{equation}
which imposes a specific relation on the set of initial 
data and is required, in particular, for the numerical integration of the
background evolution.

During the post big-bang phase, the first order perturbation of Eqs.
(\ref{g1})--(\ref{g3})  provides the linear (coupled) system of evolution
equations of the inhomogeneities. To first order in the scalar
metric fluctuations, the  line element (\ref{lel}) can be written as
\cite{mfb} \begin{equation}
ds^2 = a^2(\eta) \{( 1 + 2 \phi) d\eta^2 - 2 \partial_{i}B dx^{i} d\eta^2 -
[ ( 1 - 2 \psi) \delta_{i j} + 2 \partial_{i} \partial_{j} E ] dx^i dx^j \}.
\label{pertlel}
\end{equation}
Since there are  two gauge transformations preserving the 
scalar nature of the above metric fluctuations $(\psi, \phi,E,B)$,  
 two gauge-invariant 
(Bardeen) potentials can be defined \cite{mfb,bar}:
\begin{eqnarray}
&&\Phi = \phi + \frac{1}{a} [ ( B- E') a]',
\label{b1}\\
&& \Psi = \psi - {\cal H} ( B - E'). 
\label{b2}
\end{eqnarray}
Appropriate gauge-invariant variables 
can also be  defined 
for the perturbations of the sources, in such a way that 
\begin{eqnarray}
&&\chi^{(\rm gi)} = \delta\sigma + \sigma' ( B - E') ,
\label{chidef}\\
&& \delta\rho^{(\rm gi)} _{\rm r} = \delta\rho_{\rm r} + 
\rho_{\rm r}' ( B - E'), 
\label{rhrdef}\\
&& v_{\rm r}^{(\rm gi)} = v_{\rm r} + (B -E'),
\label{potdef}
\end{eqnarray}
whose physical interpretation is particularly simple  in the
so-called longitudinal gauge \cite{mfb} in which $E=0=B$. Here 
$\delta {\cal T}_{0}^{0} = \delta \rho_{\rm r}$, and the  velocity
potential is defined by the off-diagonal fluctuations  of the radiation
energy-momentum tensor as  \begin{equation}
\delta {\cal T}_{i}^{0} = (p_{\rm r} + \rho_{\rm r}) u^{0} \delta u_{i},
\end{equation}
where $ u^{0} = 1/a$ and, in the longitudinal gauge, 
$ \delta u_{i} = a \partial_{i} v_{\rm r}$. 

By perturbing the diagonal components of $T_{\mu}^{\nu}(\sigma)$, and
using 
 Eqs. (\ref{b1}) and (\ref{chidef}),
the fluctuations of the axionic energy and pressure 
densities can be expressed in a fully gauge-invariant way as follows 
\footnote{ In the 
following, since we will be dealing only with gauge-invariant quantities, 
 the superscript ``$({\rm gi})$'' can be consistently dropped 
without confusion. }: 
\begin{eqnarray}
&& \delta \rho_{\sigma} = \frac{1}{a^2} \biggl[ - \Phi {\sigma'}^2 + \sigma' 
\chi' + \frac{\partial V}{\partial\sigma} a^2\chi \biggr],
\label{deltarhosigma}\\
&& \delta p_{\sigma} = \frac{1}{a^2} \biggl[ - \Phi {\sigma'}^2 + 
\sigma'  \chi' - \frac{\partial V}{\partial\sigma} a^2 \chi\biggr].
\label{deltapsigma}
\end{eqnarray}
The  variables characterizing the gauge-invariant 
fluctuations of the sources can be defined in different, but equivalent, 
ways \cite{ks,dur}. 
For instance, it is sometimes useful (especially in the case 
of fluids with constant speed of sound) to write 
equations for the combination $(\delta \r_r/\r_r- 4 \Phi)$,  whose 
evolution greatly simplifies at large scales.

The fluctuations of the off-diagonal (space-like) components of  Eq.
(\ref{g1}) imply that $\Phi = \Psi$. Hence, in terms of the variables
defined in Eqs.  (\ref{b1})--(\ref{deltapsigma}), the ($00$) and ($0i$)
components of the perturbed Einstein equations (acting as Hamiltonian
and momentum constraints   for the evolution of the Bardeen potential)
can be written in terms of the gauge-invariant velocity potentials 
$v_{\rm r}$, $v_{\sigma}$, and of   the radiation and axion density
contrasts $ \delta_{\rm r}=\delta\rho_{\rm r}/\rho_{\rm r}$,  
$\delta_{\sigma} = \delta \rho_{\sigma}/\rho_{\sigma} $,  as follows: 
\begin{eqnarray}
&& \nabla^2 \Phi - 3 {\cal H} ( {\cal H} \Phi + \Phi') = 
\frac{a^2}{4} \biggl( \rho_{{\rm r}} \delta_{\rm r} + \rho_{\sigma} 
\delta_{\sigma} \biggr),
\label{hamp}\\
&& {\cal H} \Phi + \Phi' = \frac{a^2}{4}\biggl[ (\rho_{\rm r} + p_{\rm r} ) 
v_{\rm r} + (\rho_{\sigma} + p_{\sigma}) v_{\sigma} \biggr],
\label{momp}
\end{eqnarray}
Here the 
axion velocity potential, $v_{\sigma}$, is defined by
\begin{equation}
v_{\sigma} = \frac{ \chi}{a \sqrt{ p_{\sigma} + \rho_{\sigma}}}
\label{usigma}
\end{equation}
and  is the axionic counterpart of
the  velocity potential introduced for the  radiation fluid.

The constraints  (\ref{hamp}) and (\ref{momp})
are to be supplemented by the dynamical equations coming from the 
perturbation of the ($ii$)  components of Einstein's
equations (\ref{g1}), of the axion equation  (\ref{g2}) and of the
continuity equation (\ref{g3}).  For the
gauge-invariant quantities defined above, such dynamical equations
are, respectively: 
\begin{eqnarray} && \Phi'' + 3 {\cal H} \Phi' + ( {\cal
H}^2 + 2 {\cal H}') \Phi =  \frac{a^2}{12} \rho_{\rm r} \delta_{\rm r} +
\frac{a^2}{4} \delta p_{\sigma} , \label{ij}\\
&& \chi'' + 2 {\cal H} \chi' - \nabla^2 \chi + 
\frac{\partial^2 V}{\partial\sigma^2} a^2 \chi - 4 \sigma' \Phi' + 2 
\frac{\partial V}{\partial \sigma }a^2 \Phi =0,
\label{chired} \\
&& \delta_{\rm r}' - 4 \Phi' - \frac{4}{3} \nabla^2 v_{\rm r} =0,
\label{deltar}\\
&& v_{\rm r}' - \frac{1}{4} \delta_{\rm r} - \Phi =0.
\label{ureq}
\end{eqnarray}
Finally, the perturbation of the covariant conservation of the axionic
energy-momentum tensor leads to two useful equations:
\begin{eqnarray}
&& \rho_{\sigma} \delta_{\sigma}' - ( p_{\sigma} + \rho_{\sigma}) 
\nabla^2 v_{\sigma} - 3 {\cal H} p_{\sigma} \delta_{\sigma} 
- 3 \Phi' ( p_{\sigma} + \rho_{\sigma} ) + 3 {\cal H} \delta p_{\sigma} =0,
\label{contsig}\\
&& v_{\sigma}' + \biggl( 4 {\cal H} + 
\frac{ p_{\sigma}' + \rho_{\sigma}'}{p_{\sigma} + 
\rho_{\sigma}}\biggr) v_{\sigma} - \frac{\delta p_{\sigma}}{p_{\sigma} + 
\rho_{\sigma} } - \Phi =0, 
\label{usigeq}
\end{eqnarray}
which are implied, as it should be,  by Eqs. (\ref{ij})--(\ref{ureq}) when 
the background equations (\ref{dyn})--(\ref{kg}) are used.

It is also useful to notice that, by combining Eqs. (\ref{ham}) and 
(\ref{ij}), we can eliminate  the fluid variables, and we obtain
\begin{equation}
\Phi'' + 4 {\cal H} \Phi' + 2 ( {\cal H}^2 + {\cal H}') \Phi - \frac{1}{3} 
\nabla^2 \Phi = - \frac{{\sigma'}^2}{6} \Phi + \frac{\sigma'}{6} \chi' 
- \frac{1}{3} \frac{\partial V}{\partial \sigma} a^2 \chi, 
\label{phred}
\end{equation}
which, together with Eq. (\ref{chired}), provides a closed
system  of equations for $\Phi$ and $\chi$. Of course, the velocity 
potential and the  density contrast of the fluid do not disappear 
from the physics of our problem,  and have to be 
directly computed using the Hamiltonian and momentum 
constraints of Eqs. (\ref{hamp}) and (\ref{momp}).

\subsection{Curvature perturbations from non adiabaticity}

Given the system of Eqs. (\ref{ij})--(\ref{ureq}), supplemented 
by the constraints (\ref{hamp})--(\ref{momp}), it is 
sometimes appropriate to select variables obeying simple 
evolution equations in the long-wavelength limit, in which 
the spatial gradients are negligible. For this purpose, 
a particular combination of Eqs. (\ref{hamp})
and  (\ref{ij}) will be considered, and 
the fluctuations in the  total energy and pressure densities will be 
defined:
\beq
\delta \rho_{\rm tot} = \delta \rho_{\sigma}+ \delta \rho_{\rm r},
~~~~~~~~~~~
 \delta p_{\rm tot} = \delta p_{\sigma}+ \delta p_{\rm r}. 
\label{tot}
\eeq 
In terms of the quantities 
defined in Eq. (\ref{tot}),
 the  evolution of the Bardeen potential can be formally 
written in terms of a single equation
\begin{equation}
\Phi'' + 3 {\cal H} ( 1 + c_{s}^2) \Phi' + 
[ 2 {\cal H}' + {\cal H}^2 ( 1 + 3 c_{s}^2)]\Phi - c_s^2 \nabla^2 \Phi= 
\frac{a^2}{4} [ \delta  p_{\rm tot} - c_{s}^2 \delta \rho_{\rm tot}],
\label{source}
\end{equation}
where $c_s$ is the speed of sound for the total system, defined by
\begin{equation}
c_{s}^2 = \frac{ p_{\rm tot}'}{\rho_{\rm tot}'} \equiv 
\frac{ p_{\sigma}' + p_{\rm r}'}{\rho_{\sigma}' + \rho_{\rm r}'}, 
\end{equation}
or, using the explicit form of the background equations, 
\begin{equation}
c_{s}^2 =
\frac{1}{3}\biggl\{ \frac{\rho_{\rm r} 
+ \frac{9}{4} ( p_{\sigma} + \rho_{\sigma}) +
\frac{3}{2} \frac{\sigma'}{{\cal H}} V_{,\sigma}     }{\rho_{\rm r} + 
\frac{3}{4} ( p_{\sigma} + \rho_{\sigma})}\biggr\}, 
\label{sps}
\end{equation}
where $V_{,\sigma} \equiv \pa V/\pa \sg$.

The left-hand side of Eq. (\ref{source}) (except for  the Laplacian term) 
can now be expressed as  the time  derivative of a single
gauge-invariant function $\zeta$, namely  
\begin{equation}
\zeta = - \biggl[ \Phi + \frac{4{\cal H}}{a^2}\biggl( \frac{ {\cal H} \Phi 
+ \Phi'}{\rho_{\rm tot} + p_{\rm tot}}
\biggr) \biggr] \equiv - \biggl( \Phi + {\cal H} \frac{ {\cal H} \Phi 
+ \Phi'}{{\cal H}^2 - {\cal H}'}\biggr),
\label{zeta1}
\end{equation} 
where the second equality follows by using the background 
equations of motion  (\ref{dyn})--(\ref{ham}). By using this variable,
Eq. (\ref{source}) can be written as
\begin{equation}
\frac{ d \zeta}{d\eta} 
 = - \frac{{\cal H}}{p_{\rm tot} + \rho_{\rm tot}} \delta p_{\rm nad} 
- \frac{ 4 {\cal H} c_{s}^2}{a^2 (\rho_{\rm tot} + p_{\rm tot})}
\nabla^2 \Phi.
\label{zetapr}
\end{equation}
where we have defined
\begin{equation}
\delta p_{\rm nad} = \delta p_{\rm tot} - c_s^2 \delta \rho_{\rm tot}. 
\label{nad}
\end{equation} 
As noticed long ago \cite{ks,bar,lyt}, the variable $\zeta$ represents 
the inhomogeneities  in the spatial part of the space-time curvature,
measured with respect to comoving hypersurfaces ($\sigma=  {\rm
constant}$). 
Using Eq. (\ref{momp}), the variable $\zeta$  can also be usefully 
related to the total velocity potential as
\begin{equation}
\zeta = - ( \Phi + {\cal H} v_{\rm tot}),
\label{zeta}
\end{equation}
where 
\begin{equation}
(p_{\rm tot} + \rho_{\rm tot}) v_{\rm tot} = ( p_{\rm r} + \rho_{\rm r}) 
v_{\rm r} + ( p_{\sigma} + \rho_{\sigma} )v_{\sigma}. 
\label{utot}
\end{equation}
In our specific case, using the full set of background and perturbation
equations in the long-wavelength limit, where  
$ \delta_{\rm r} \sim 4 \Phi$ according to Eq. (\ref{deltar}), 
the expression for $\zeta$ can be written in the following convenient
form  
\begin{equation}
\zeta= - \frac{ \frac{3}{4} (p _{\sigma} + \rho_{\sigma}) \delta_{\rm r} 
- \rho_{\sigma} \delta_{\sigma} }{4 \rho_{\rm r} + 3 
( p_{\sigma} + \rho_{\sigma}) }. 
\label{zetaex}
\end{equation}

As we will discuss in detail in Sect. V,  in the absence of a dominant 
radiation fluid $\delta p_{\rm nad}$ is zero  at
large scales, i.e.  up to terms containing the Laplacian of $\Phi$.
However, in a radiation dominated regime, 
 $\delta p_{\rm nad} \neq 0$ and Eq.
(\ref{zetapr})  implies $\zeta' \neq 0$ even in the long-wavelength
limit. Let us then compute the
general form  of $\delta p_{\rm nad}$, for the full system of axion plus
fluid perturbations. By using the previous definitions we obtain  
\begin{equation} \delta p_{\rm nad} = \rho_{\rm r} \biggl(\frac{1}{3} -
c_{s}^2\biggr)  \delta_{\rm r}
+ \Phi ( c_{s}^2 -1) ( p_{\sigma} + \rho_{\sigma}) + 
\frac{\sigma' \chi'}{a^2} ( 1 - c_{s}^2) -
 \frac{\partial V}{\partial\sigma} \chi ( 1 + c_s^2).
\label{dpnadex}
\end{equation} 
On the other hand, using  Eqs. (\ref{hamp})
 and (\ref{momp}), we can write
\begin{equation}
\Phi(p_{\sigma} + \rho_{\sigma}) = - \frac{4}{a^2} \nabla^2 \Phi
+ 3 {\cal H} \frac{ \sigma' \chi}{a^2} + 4 {\cal H} \rho_{\rm r} v_{\rm r}
+ \rho_{\rm r} \delta_{\rm r} +
 \biggl[ \frac{\sigma' \chi'}{a^2} + \frac{\partial V}{\partial\sigma}
 \chi\biggr].
\label{intermed}
\end{equation}
Thus (neglecting 
the spatial gradient of $\Phi$) we get 
\begin{equation}
\delta p_{\rm nad} = - \frac{2}{3} \rho_{\rm r} \delta_{\rm r} 
- 2 \frac{\partial V}{\partial\sigma} \chi  + 4 {\cal H} \rho_{\rm r} 
v_{\rm r} 
(c_s^2 -1) 
+3 {\cal H} (c_s^2 -1) \frac{ \sigma' \chi}{a^2}. 
\label{dpnad2}
\end{equation}
The above equations are useful to compute, in some specific phase  of
the dynamical evolution, the source term of Eq. (\ref{zetapr})  whose
integration   allows us to obtain the explicit time  dependence of $\zeta$. 
In  \cite{us} we have determined the evolution of the fluctuations by 
following the $\zeta$ variable. In the present investigation  we will
solve the perturbation equations both in terms  of $\Phi$ and
$\zeta$,  checking numerically  the consistency of the two
approaches.

\renewcommand{\theequation}{3.\arabic{equation}}
\setcounter{equation}{0}
\section{Post big-bang histories}

At the beginning of the post-big bang evolution the background is
characterized by a ``maximal" curvature scale $H_1$, whose finite
value regularizes the big bang singularity of the standard cosmological
scenario, and provides a natural cutoff for the spectrum of
quantum fluctuations amplified by the phase of pre-big bang inflation
(see below, in particular Section VI).  In string cosmology models such
an initial curvature scale is at most of the order of the string mass
scale, i.e. $H_1 \laq M_{\rm s} \sim 10^{17}$ GeV. 

The Kalb--Ramond axion
has gravitational coupling to photons and to the QCD 
topological current but it is not necessarily identified 
with the invisible axion \cite{kim} usually 
invoked in the explanation of the strong CP problem via 
an initial misalignment of the QCD vacuum angle $\vartheta$ \cite{pre}.
The potential of Kalb--Ramond axion is, strictly speaking, 
periodic. The periodicity of the potential occurs whenever 
a Peccei--Quinn symmetry is spontaneously broken down to 
a discrete symmetry corresponding to shifting the 
$\vartheta$ angle by multiples 
of $2\pi$  (see, for instance, \cite{dvv}).
However, close to the 
minimum of the potential (i.e. sufficiently late in the process of 
 relaxation) the potential can  be assumed to be quadratic.  
Such an approximation is expected to
 be realistic for values of $\sigma$ that are small compared to its
periodicity.
Unfortunately, translating periodicity in $\vartheta$ into periodicity in $\sigma$
involves a normalization factor that is unknown in the strong-coupling region
where the dilaton is supposed to be frozen at 
late times. For this reason,  we shall keep 
the initial dispacement in Planck units, 
$\sigma_{\rm i}$, as a free parameter.

We start our study of the background and perturbation evolution at
an  initial curvature scale $H_{\rm i}\leq H_1$, when the energy density
of the  background is  mainly stored in the radiation fluid,  while the
energy density of the axion is dominated by the potential:
\begin{equation} \rho_{\rm r}(\eta_{\rm i}) \gg
\rho_{\sigma}(\eta_{\rm i})\simeq  V(\eta_{\rm i}).
\end{equation}
During the first stages of the evolution $\sigma$ remains
approximately fixed at the initial  value $\sigma_{\rm i}$ up to
corrections  ${\cal O}(V_{,\sigma})$. In the course of such a 
``slow-roll"  phase, the curvature scale  of the background decreases, 
until it becomes comparable with the curvature of the potential.
The axion background  will then start oscillating, 
at a typical scale 
\begin{equation}
H_{\rm osc} \sim m  
\label{osc}
\end{equation}
(note that, as already mentioned, we are assuming 
that the potential is quadratic).
At the curvature  scale 
\begin{equation}
H_{\sigma} \sim m \sigma(t),
\label{dominance}
\end{equation}
the axion field will dominate the background. 
The specific value of the scale $H_{\sigma}$ depends upon $\sigma_{\rm i}$ 
and also upon the evolution after $\eta_{\rm i}$.
In fact,  during the oscillatory phase the axionic energy density 
decreases, on the average, as $a^{-3}$, i.e. slower than the energy of 
the radiation background, $\rho_{\rm r}
\simeq a^{-4}$. 
From Eqs. (\ref{osc}) and (\ref{dominance}) 
it is then  clear  
that, depending on the initial value of $\sigma$, the
oscillations of the axionic background may arise either before 
or after the phase of $\sigma$-dominance. 

Irrespectively of its initial value,  
the coupling of $\sigma$ to photons is gravitational, i.e. 
suppressed by the Planck mass. 
The decay takes place when the curvature 
scale is of the same order as the decay rate, namely when
\begin{equation}
H\sim H_{\rm d} \sim  \frac{m^3}{M_{\rm P}^2}.
\label{decb}
\end{equation}
The late decay of $\sigma$ is in general associated to a
significant entropy release, which has to be carefully
constrained \cite{ell1,ell2,g3} not to spoil the light nuclei abundances
and the baryon asymmetry generated, respectively, by
primordial nucleosynthesis and baryogenesis.

In our context, for typical values of $H_1$, and for a
realistic scenario, the decay of $\sigma$ is constrained to occur
prior to nucleosynthesis, i.e. at a scale 
$H_{\rm d}>H_{\rm N} \sim
(1 {\rm MeV})^2/M_{\rm P}$, which implies $m \gaq 10$ TeV.
The lower bound on the axion mass is even larger if we
require that the decay occurs prior to baryogenesis at the
electroweak scale (characterized by a temperature of the
cosmological plasma of order $0.1$ TeV), which implies
$m \gaq 10^4$ TeV. If, on the contrary, baryogenesis occurs at a
large enough scale preceeding the phase of axion
dominance and decay, then the minimal value of $m$
allowed by the entropy constraints \cite{ell1,ell2,g3} is, in general,
$\sg_{\rm i}$-dependent. In that case, however, the resulting
lower bound is strongly dependent on the given model of
baryogenesis, and can be somewhat relaxed by various
mechanisms. In the rest of this paper we will thus adopt a
conservative approach, by taking the nucleosynthesis bound
$m \gaq 10^{-14} M_{\rm P}$ as a typical reference value.

\subsection{Late dominance of the axion: $\sigma_{\rm i} <1$}

If $\sigma(\eta_{\rm i}) = \sigma_{\rm i} < 1$,
then 
the axionic background  first experiences 
a phase of radiation-dominated oscillations,
from $H_{\rm osc}$ down to $H_{\sigma}$.  
The duration of this phase depends upon $\sigma_{\rm i}$, since 
$(a_{\rm osc}/a_{\sigma}) \sim \sigma_{\rm i}^2$, and it may be rather
long,   if $\sigma_{\rm i} \ll 1$. 
During this phase the axion potential energy decreases as
$a^{-3}$.  Consequently, 
the typical scale of axion dominance is, from Eq. (\ref{dominance}), 
\begin{equation}
H_{\sigma} \simeq m \sigma_{\rm i}^4.
\end{equation}
From $H_{\sigma}$ down to $H_{\rm d}$, i.e. inside the regime 
of axion-dominated oscillations, 
the effective equation of state of the gravitational sources, averaged 
over one  oscillation, mimics that of 
dusty matter, with  $\langle p_{\sigma} \rangle =0$.
During this regime the scale factor and the Hubble 
parameter also have oscillating corrections, which vanish on the
average, and decay away for large  times. It should be stressed,
however, that  the effective equation of state of the axion background,
for curvature scales  smaller than $H_{\sigma}$,
depends upon the curvature of the potential 
around the minimum. If, for instance, the potential 
is not quadratic, but quartic, the coherent 
oscillations will lead to an effective equation of state 
that simulates a radiation fluid, i.e.
 $3 \langle p_{\sigma}\rangle = \langle \rho_{\sigma}\rangle$ \cite{tur}.

The occurrence of the axion-dominated phase requires
\begin{equation} 
H_{\rm d} < H_{\sigma},
\label{h1}
\end{equation}
which imposes a lower bound on the 
initial axionic amplitude, namely
\begin{equation}
1>\sigma_{\rm i} > \sqrt{{m}/{M_{P}}}\equiv \bar{\sigma}.
\label{c1}
\end{equation} 
This constraint, 
however,  is not so demanding, given the generous lower
bound on $m$ (in Planck units) allowed by nucleosynthesis and
baryogenesis. Finally, after the axion decay, the Universe enters a
subsequent  
radiation-dominated epoch. From this moment on, the evolution 
of the background fields is standard.  

\subsection{Early dominance of the axion: $\sigma_{\rm i} >1$} 

If $\sigma(\eta_{\rm i}) =\sigma_{\rm i} >1$, 
then the axion, right after 
the onset of the radiation-dominated epoch, starts again rolling down
its potential. This initial part of the 
evolution is completely analogous to that of the $\sigma_{\rm i} < 1$ 
case. 
However, for $\sigma_{\rm i} >1$, the axion dominance 
will occur before the onset of the axion oscillating phase,
i.e.
\begin{equation}
H_{\rm osc} < H_{\sigma},  
\label{h2}
\end{equation}
where, for a generic potential, 
\begin{equation}
H_{\sigma} \sim \sqrt{V(\sigma_{\rm i})}
\label{dom2}
\end{equation} 
(since the kinetic energy of the axion 
is negligible during the  slow-roll evolution).  
At $H=H_{\sigma}$ the Universe enters a phase of accelerated
expansion (slow-roll inflation) whose duration, for a quadratic
potential, is given by  
\begin{equation}
Z_{\sigma} = \frac{a_{\rm  final}}{a_{\rm initial}}= 
\exp\left[{ \frac{1}{8} ( \sigma_{\rm initial}^2 - \sigma_{\rm
final}^2)}\right].  
\label{srdurtaion}
\end{equation}
This
inflationary phase will last until $H=H_{\rm osc}\sim m$, 
$\sigma_{\rm final}
 \simeq 1$
(if we assume, again,  that close to its minimum the potential is 
quadratic).  For $H<H_{\rm osc}$ the background will be dominated by
the coherent  oscillations of the axion, whose decay will eventually
produce a second radiation-dominated phase (in full analogy with the
case  $\sigma_{\rm i} <1$). 

This scenario requires, for consistency, that 
\begin{equation}
H_{\sigma} < H_{\rm i}\leq H_1 \leq M_{\rm s}, 
\end{equation}
which, in the case 
of a quadratic  potential, amounts to require 
\begin{equation}
{H_{1}}/{m}\geq {H_{\rm i}}/{m} \gaq \sigma_{\rm i} > 1.
\label{c2}
\end{equation}
As in the case of Eq. (\ref{c1}), this bound is not so restrictive,
given the limits on the axion mass. Indeed, in the case 
$\sigma_{\rm i} >1$, the most stringent constraints are  coming 
not from the 
evolution of the background geometry but, as we shall see, from the
evolution  of the fluctuations that forbid too large
values of $\sigma_{\rm i}$. One is then left with a situation where 
$\sigma_{\rm i} \simeq 1$ and $H_{\sigma} \simeq H_{\rm osc}$. In
such a case, the phase of  axion-dominated oscillations will take place
right after the radiation-dominated period of slow-roll, without
 a long intermediate epoch of inflation.

\subsection{Initial conditions for the fluctuations}

Given the coupled system of gauge-invariant perturbation equations,
the initial conditions for  the Bardeen potential, for the perturbed
radiation  density and for the radiation velocity field, will be imposed as
follows
 \begin{equation}
\Phi_{k}(\eta_{\rm i}) =0,~~~~~\delta_{\rm r}(\eta_{\rm i}, k) =0, 
~~~~v_{\rm r}(\eta_{\rm i},k) =0, 
\label{incon3}
\end{equation}
assuming that no appreciable amount of adiabatic metric perturbations
has been directly generated (on large scales) by the 
 pre-big bang dynamics. The only non-vanishing initial 
fluctuations are the (isocurvature) axionic seeds, amplified from
the vacuum during the  pre-big bang evolution: 
\begin{equation}
\chi_{k}(\eta_{\rm i}) = \chi_{\rm i} (k)\neq 0,
\end{equation}
so that,  from Eq. (\ref{deltarhosigma}),
$\delta_{\sigma} (\eta_{\rm i}, \vec{x} ) \neq 0$.

In the present analysis we shall assume that the amplitude of the
initial  axion fluctuations is smaller (for all modes) 
than $\sigma_{\rm i}$, i.e. 
\begin{equation}
k^{3/2}|\chi_{\rm i}(k)| < \sigma_{\rm i}.
\label{incon3a}
\end{equation}
In the opposite case, $k^{3/2}|\chi_{\rm i}(k)| > \sigma_{\rm i}$, 
we are led to the case already analysed in 
\cite{dgs1,dgs2,dgs3,mvdv} where $\sigma_{\rm i}$ was assumed 
to be zero, and the obtained 
large-scale fluctuations are of the isocurvature 
type, and strongly non-Gaussian. 
If $\sigma_{\rm i} >\bar{\sigma}$, the non-Gaussianity is  rather
small,  but can be enhanced  if the axion does not dominate 
at decay ($\sigma_{\rm i} < \bar{\sigma}$) \cite{lw,luw}.

\renewcommand{\theequation}{4.\arabic{equation}}
\setcounter{equation}{0}
\section{Background and perturbation evolution for
$\sigma_{\i} < 1$}
 
In view  
of the forthcoming phenomenological applications,
the main quantity that we need to compute is the 
spectral amplitude of the Bardeen potential after axion 
decay, during the subsequent radiation-dominated evolution, as a 
function of  the  spectral
amplitude of the axion fluctuations amplified by the phase of  pre-big
bang inflation. It is important, for this purpose, to have a reasonably
accurate control on the evolution of the background and of the 
fluctuations. Using different approximations, motivated  by the
hierarchy of scales discussed in the previous section, we will first
analytically determine  the evolution 
of the system through the different  cosmological stages. 
Numerical integrations will then be used  in order to check  
the analytical results in  the cross-over
regimes connecting the different phases of the background evolution.

\subsection{The radiation-dominated  slow-roll regime}

During the  first stage of evolution, 
  $\rho_{\rm r}(\eta_{\rm i}) \gg V(\eta_{\rm i})$. In this limit, Eqs. 
(\ref{dyn})--(\ref{contrad})  and (\ref{ham}) simplify
to:
\begin{eqnarray}
&&{\cal H}^2 = \frac{a^2}{6} \rho_{\rm r},
\label{slowrolla}\\
&& \rho_{\rm r}' + 4 {\cal H} \rho_{\rm r} =0,
\label{slowrollb}\\
&& \sigma'' + 2 {\cal H} \sigma' + a^2 \frac{\partial V}{\partial\sigma} 
=0.
\label{slowroll}
\end{eqnarray}
Equations (\ref{slowrolla}) and (\ref{slowrollb})  
imply that ${\cal H} a$ is constant. Furthermore, 
since the kinetic energy of $\sigma$ is subleading with respect to
 the potential, the axionic field 
slowly rolls down the potential.
In such a situation a systematic expansion 
in the gradient of the potential,  $V_{,\sigma}$, 
can be developed, and the background evolution
 is adequately described by the following  approximate 
equation
\begin{equation}
\sigma' = - \frac{1}{5} \frac{a^2}{\cal H} V_{,\sigma},
\label{sre}
\end{equation}
which can be integrated to give
\begin{equation}
\sigma \simeq \sigma_{\rm i} - \frac{1}{20}
\biggl(\frac{V_{,\sigma}}{H^2}- 
\left. \frac{V_{,\sigma}}{H^2} \right|_{\rm i}\biggr), 
\label{sre2}
\end{equation}
i.e. $\sigma$  is approximately constant up to 
corrections that depend  upon the specific form of the 
potential, and which induce a slight decrease of the 
axion background.

In order to solve 
the Hamiltonian constraint (\ref{hamp})
it is now convenient to work in terms of the  Fourier components of the
perturbation variables, $\Phi_k, \delta_{\rm r}(k)$, and so on.  Since we are
interested in large scale  inhomogeneities we first obtain,  
 from Eq. (\ref{deltar}),
\begin{equation}
\delta_{\rm r} (k) \simeq 4 \Phi_{k} ,
\label{drph}
\end{equation} 
where the integration constants vanish because of Eq. (\ref{incon3}).
Consequently, using Eq. (\ref{deltarhosigma}),   the Hamiltonian
constraint (\ref{hamp}) can be written as  
\begin{equation}
- 3 {\cal H} ( {\cal H} \Phi_{k} + \Phi_{k}' )  - 
\Phi_{k}\biggl[ a^2 \rho_{\rm r} - \frac{{\sigma '}^2}{4}\biggr] 
= \frac{1}{4} \biggl[ \sigma' \chi_{k}' + V_{,\sigma}
a^2 \chi_{k}\biggr],
\label{hamint}
\end{equation}
where the spatial gradients have been consistently 
neglected.
Using Eq. (\ref{slowrolla}), 
\begin{equation}
\Phi_{k}' + 3 {\cal H}\Phi_{k} 
\simeq - \frac{1}{12 {\cal H} }\biggl[ \sigma' \chi_{k}' 
+ \frac{\partial V}{\partial\sigma} a^2 \chi_{k}\biggr].
\label{app}
\end{equation}
On the other hand, from Eq. (\ref{chired}), the evolution of $\chi_{k}$
is approximately given by  
\begin{equation}
\chi_{k}' \simeq - \frac{1}{5} V_{,\sigma\sigma} 
\frac{a^2}{{\cal H}} \chi_{k}.
\label{chev}
\end{equation} 
The first term  on the r.h.s. of Eq. (\ref{app}) thus 
contains three derivatives of the potential, and it is 
subleading with respect to the second term. 
Direct integration of Eq. (\ref{app})  then gives 
\begin{equation}
\Phi_{k}(\eta) = - \frac{1}{84} \frac{a^2}{{\cal H}^2} 
V_{,\sigma} \chi_{k} + {\cal O}( V_{,\sigma}^2) \simeq 
- \frac{1}{14~\rho_{\rm r}} V_{,\sigma} \chi_{k} 
+ {\cal O}( V_{,\sigma}^2). 
\label{phgen}
\end{equation}  
As a consequence, from Eqs. (\ref{deltar}) and (\ref{ureq}) we can
determine $\delta_{\rm r}$ and  $v_{\rm r}$ as
\beq
 \delta_{\rm r}(k,\eta) = - \frac{1}{21} \frac{a^2}{{\cal H}^2} 
V_{,\sigma} \chi_{k} + {\cal O}( V_{,\sigma}^2),
~~~~
 v_{\rm r}(k,\eta) = - \frac{1}{210} \frac{a^2}{{\cal H}^3} 
V_{,\sigma} \chi_{k} + {\cal O}( V_{,\sigma}^2).
\label{flsol}
\eeq
Inserting now the obtained solutions in the remaining equations
 (\ref{momp}) and (\ref{ij})
 (not used for the above derivation), we can see that they are satisfied
with the same accuracy.

The time evolution
of $\zeta_{k}$ in the radiation-dominated, slow-roll regime can 
finally be determined from Eq. (\ref{zeta1}):
\begin{equation}
\zeta_{k}(\eta) \simeq \frac{1}{4 \rho_{\rm r}} 
\frac{\partial V}{\partial\sigma}
\chi_{k} + {\cal O}( V_{,\sigma}^2),
\label{zet2}
\end{equation}
so that $\Phi_{k}$ and $\zeta_{k}$ obey the
following  simple relation
\begin{equation}
\Phi_{k}(\eta) \simeq -(2/7) \zeta_{k}(\eta) + {\cal O}( V_{,\sigma}^2).
\end{equation}

It should be appreciated  that  Eq. 
 (\ref{zet2}) can also be obtained by direct integration of  Eq. 
(\ref{zetapr}).
In  the limit $(p_{\sigma} + \rho_{\sigma}) \ll \rho_{\rm r}$, 
Eq. (\ref{sps}) implies indeed
\begin{equation}
c_{s}^2 \simeq \frac{1}{3} + \frac{1}{2\rho_{\rm r}}
 \frac{\sigma'}{{\cal H}} V_{,\sigma}.
\label{sps1}
\end{equation}
On the other hand, from  Eqs. (\ref{dpnadex}) and  
(\ref{drph}), 
 the approximate  form of $\delta p_{\rm nad}(k)$ is 
\begin{equation}
\delta p_{\rm nad} (k) \simeq - \frac{4}{3} V_{,\sigma} \chi_{k} + 
{\cal O}( V_{,\sigma}^2) 
\label{dpnadsr}
\end{equation}
(again, terms with more than one derivative of the potential 
have been neglected). By 
inserting this result into Eq. (\ref{zetapr}) 
we are led to the equation 
\begin{equation}
\frac{d \zeta_{k}}{d \ln{a}} = \frac{V_{,\sigma} \chi_{k}}{\rho_{\rm r}},
\end{equation}
whose direct integration  (recall that $\rho_{\rm r} \sim a^{-4}$) 
leads, as expected, to Eq. (\ref{zet2}).

The above approximate results for $\Phi_{k}$ and $\zeta_{k}$
hold for a generic (flat enough) 
potential.  However, in order to check  the correctness of
our approximations numerically, it is useful to consider the
simple case  of a quadratic potential:
\begin{equation}
V(\sigma) = \frac{m^2}{2} \sigma^2.
\end{equation}
In that case,  Eqs. (\ref{sre}), 
(\ref{chev}), (\ref{phgen}), (\ref{zet2}) lead to  
\begin{eqnarray}
&& \sigma(\tau) \simeq \sigma_{\rm i} \left[ 1 - \frac{\mu^2}{20}
\left(\tau^4 -1\right) + {\cal O}( \mu^4 \tau^8) \right],
\label{sigsr}\\
&& \chi_{k}(\tau) \simeq \chi_{\rm i}(k) \left[ 1 - \frac{\mu^2}{20}
\left(\tau^4 -1\right)  + {\cal O}( \mu^4 \tau^8) \right],
\label{chsr}\\
&&
\Phi_{k}(\tau) \simeq  - \frac{\sigma_{\rm i} \chi_{\rm i}(k)}{84}\left[
\mu^2  \left(\tau^4 -1\right) + {\cal O}( \mu^4 \tau^8)\right],
\label{phsr}\\
&&\zeta(\tau) \simeq
\frac{\sigma_{\rm i} \chi_{\rm i}(k) }{24 } \left[\mu^2 
\left(\tau^4 -1\right) 
+ {\cal O}( \mu^4 \tau^8)\right],
\label{zetsr}
\end{eqnarray}
where $\chi_{\rm i}(k)$ is  the initial spectrum of the axionic 
fluctuations, and we have defined  the  (dimensionless) rescaled mass
and conformal time coordinate:   
\begin{equation}
\tau= \frac{\eta}{\eta_{\rm i}}, ~~~~~\mu = m\,\eta_{\rm i}\,a_{\rm i}=m/H_{\rm i}. 
\label{resc}
\end{equation} 
The time  $\eta_{\rm i}$ is the initial 
integration time and $a_{\rm i}$ the initial normalization 
of the scale factor.  
These rescalings are useful in order  to compare the numerical results
with  the analytical calculations.

We have performed a numerical integration by choosing 
initial conditions at sub-Planckian curvature scales, i.e. 
\begin{equation}
H_{\rm i} = \frac{{\cal H}_{\rm i}}{a_{\rm i}} \ll 1, ~~~~~~~\eta_{\rm i} \gg 1 
\end{equation}
(in Planck units), and setting  $a_{\rm i} =1$.
Given a  value of $\sigma_{\rm i}$ compatible, for a given mass, 
with Eq. (\ref{c1}), the constraint (\ref{ham})
fixes  
the initial  radiation background
$\rho_{\rm r}(\eta_{\rm i}) $ to a value  
much larger than the axionic potential. 
Similarly, 
the  initial values of the derivatives of $\Phi_{k}$ and $\chi_{k}$ are 
obtained by imposing, on the initial data 
(\ref{incon3})--(\ref{incon3a}),  the Hamiltonian 
and momentum constraints of Eqs. (\ref{hamp}) and (\ref{momp}).
It has been checked 
that  all the constraints (both for the background and for the 
fluctuations) are satisfied at every time 
all along the numerical integration. The system describing 
the evolution of the fluctuations, in particular, can be integrated 
in two different (and complementary) ways. We could either use 
Eqs. (\ref{ij})--(\ref{ureq}) and follow the evolution of all variables,
or use Eqs. (\ref{chired})--(\ref{phred}) and integrate the system 
only in terms of $\Phi_{k}$ and $\chi_{k}$. We have performed the
numerical  integration in both ways, and  checked that the results are
the same. 

\begin{figure}
\centerline{\epsfxsize = 12 cm  \epsffile{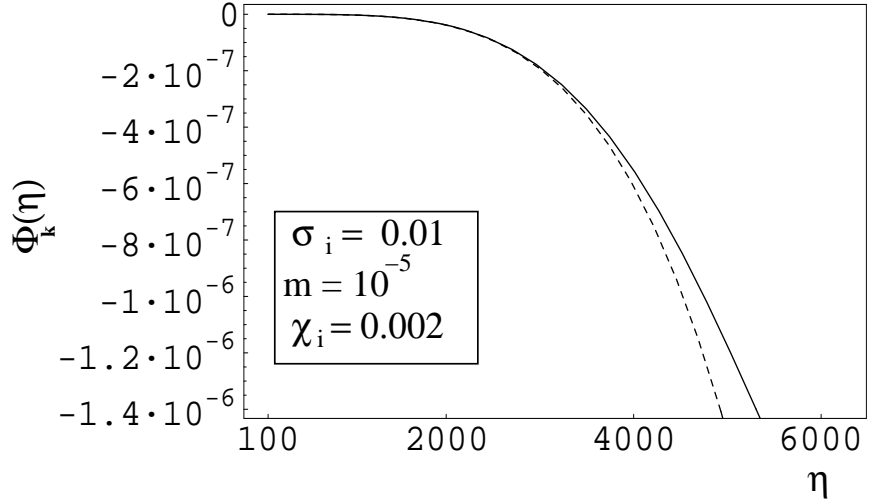}}
\vskip 3mm
\caption[a]{The full curve shows the result of a numerical 
integration for 
the case  ${\cal H}_{\rm i} = 0.01$, and for the set of parameters 
reported in the figure. The dashed curve shows the approximate 
analytical solution  based on  Eqs. (\ref{phgen}) and (\ref{phsr}).}
\label{SR1} 
\end{figure}

\begin{figure}
\centerline{\epsfxsize = 11 cm  \epsffile{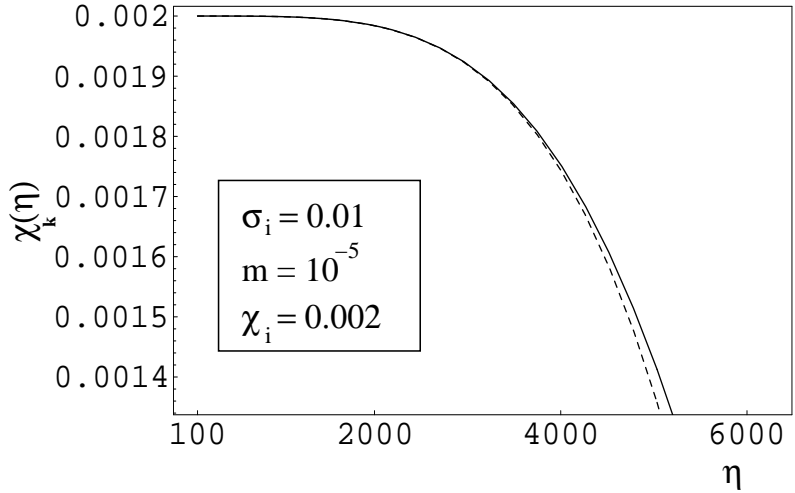}}
\vskip 3mm
\caption[a]{Evolution of $\chi_{k}$,  reported for the same 
set of parameters as in Fig. \ref{SR1}. The dashed curve 
corresponds to the approximate analytical result
of Eq. (\ref{chsr}). } 
\label{SR2}
\end{figure}

\begin{figure}
\centerline{\epsfxsize = 11 cm  \epsffile{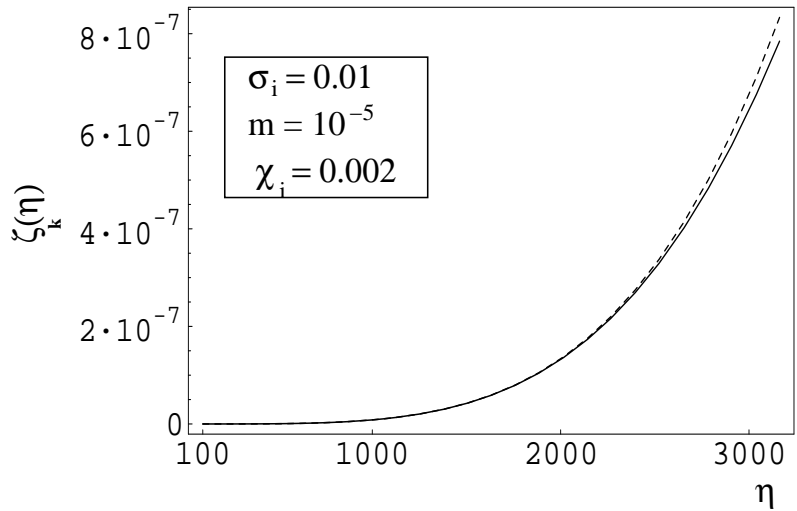}}
\vskip 3mm
\caption[a]{Evolution of $\zeta_k$, for the same set of parameters as in
Figs. \ref{SR1} and \ref{SR2}. The numerical result (full curve) is  
compared  with the approximate analytical result of Eq. (\ref{zetsr})
(dashed curve).} 
\label{SR2a} 
\end{figure}

In Figs. \ref{SR1}, \ref{SR2} and \ref{SR2a} we report, as  full
curves,  the results of the numerical integration for a quadratic
potential.  The analytical results of 
Eqs. (\ref{sigsr})--(\ref{zetsr}),  based on the slow-roll
approximation, are also illustrated, for comparison, by the dashed
curves. 

\subsection{Radiation-dominated oscillations}

During the radiation-dominated regime, and for a quadratic potential,
the axion evolution equation (\ref{kg}) can be written as
\begin{equation}
\frac{ d^2 g}{d \tau^2} + \mu^2 \tau^2 g = 0, ~~~~~g = \sigma a,
\label{f}
\end{equation}
and its exact solution can be given in terms of Bessel functions
\cite{abr} as  
\begin{equation}
g(\tau) = \sqrt{\tau} {\cal C}_{1/4} \bigg( \frac{\mu \tau^2}{2}\biggr),
\end{equation}
where ${\cal C}_{1/4}$ is a linear combination (with 
constant coefficients) of Bessel functions of 
order $1/4$ and $\mu\tau^2/2\sim m(t -t_{\rm i})$. 
By imposing the correct boundary conditions and
normalization,  in such a way that
$\sigma(\tau) \to  \sigma_{\rm i}$ for $ \tau \to 1$, we obtain
\begin{equation} 
\sigma(\tau) = \frac{\sigma_{\rm i} }{\sqrt{\tau}}
\frac{1}{J_{1/4} (\mu/2) } J_{1/4}\biggl( \frac{\mu}{2} \tau^2\biggr),
~~~~ \eta_{\rm i}<\eta<\eta_\sg, 
\label{exsig}
\end{equation}
where $J_{1/4}$ is the first-kind Bessel function. 
Notice that the small argument limit of this equation, for $\mu\ll 1 $
and  $\tau \ra 1$, exactly gives the result (\ref{sigsr}), obtained in
the  slow-roll approximation. This exact
analytical solution can also be
used  as a consistency check of the quadratic approximations 
when the  potential, during slow-roll,  has a more complicated
analytical form.
 
The onset of the axion oscillations can be determined
from the first zero of $J_{1/4}(\mu \tau^2/2)$, which 
occurs for 
\begin{equation}
\frac{\mu \tau^2}{2} \simeq 2.78,
\end{equation}
namely for 
\begin{equation}
\tau_{\rm m} = \frac{\epsilon_1}{\sqrt{\mu}}, ~~~~~\epsilon_{1} 
\simeq 2.35. 
\label{430}
\end{equation}
Different definitions of the oscillation starting time, for instance 
related to the breakdown of the slow-roll approximation, 
would lead to similar numerical values, i.e. to Eq. (\ref{430})
with $\epsilon_1 \simeq (12)^{1/4}$.  In the large argument limit
($\mu/2 \tau^2 \gg 1$)  of Eq. (\ref{exsig}) the solution finally describes
the oscillating regime, 
\begin{equation}
\sigma(\tau) \simeq \frac{ 2 \sigma_{\rm i} }{\tau^{3/2} 
\sqrt{\pi \mu} 
J_{1/4}(\mu/2) } \cos{\biggl( \frac{\mu \tau^2}{2} - 
\frac{3}{8} \pi\biggr)}, ~~~~ \eta_{\rm osc} <\eta<\eta_\sg, 
\label{wkbsig}
\end{equation}
where the phase and amplitude of oscillations are fixed by the 
initial conditions.

An approximate solution of Eq. (\ref{chired}), similar to Eq.
(\ref{exsig}),  holds for the axion fluctuations, namely
\begin{equation}
\chi_{k}(\tau) \simeq 
\frac{\chi_{\rm i}(k) }{\sqrt{\tau}} \frac{1}{J_{1/4}
(\mu/2) } J_{1/4}\biggl( \frac{\mu}{2} \tau^2\biggr),
~~~~~~ \eta_{\rm i}< \eta<\eta_\sg, 
\label{exch}
\end{equation}
whose small and large argument limits lead, respectively, to 
\begin{eqnarray}
&&   \chi_{k}(\tau) \simeq \chi_{\rm i}(k) \left[ 1 - \frac{\mu^2}{20} 
\left(\tau^4  -1\right)\right],
\label{chsmall}\\
&&\chi_{k}(\tau) \simeq \frac{ 2 \chi_{\rm i}(k) }{ \tau^{3/2} 
\sqrt{\pi \mu} 
J_{1/4}(\mu/2) } \cos{\biggl( \frac{\mu \tau^2}{2} - 
\frac{3}{8} \pi\biggr)}.
\label{chlarge}
\end{eqnarray}
We notice that Eq. (\ref{exch}) 
 is obtained by solving the {\em approximate} 
evolution equation 
\begin{equation}
\chi_{k}'' + 
2 {\cal H} \chi_{k}' + m^2 a^2 \chi_{k} 
\simeq 0, 
\label{approx}
\end{equation}
i.e. neglecting the terms containing the Bardeen potentials in 
Eq. (\ref{chired}).
 In the slow-roll approximation, as previously stressed,
these terms can be neglected for a {\em generic} potential term. 
However,  they can also be neglected  in the oscillating phase, 
provided the potential is well approximated by a  quadratic form. 
We have explicitly checked that the exact analytical solutions 
(\ref{exsig})  and (\ref{exch}) are in perfect agreement with the results
of a numerical integration performed with a quadratic potential.

Thus, for a potential which is generic during the slow-roll phase (but 
still quadratic during the oscillating regime) it will  be sufficient to
work out the  slow-roll solutions specific to that potential, from Eqs.
(\ref{sre}), (\ref{chev}), and match them (with their first 
derivative) to the WKB solutions of Eqs. (\ref{f}) and (\ref{approx}),
namely  \begin{eqnarray}
&&\sigma(\tau) \simeq \frac{\sigma_{2}}{\sqrt{2 a_{\rm i}^2 \mu \tau^3}} 
\cos{\biggl(
\frac{\mu \tau^2}{2} + \beta\biggr)},
\label{solsig2}\\
&& 
\chi_{k}(\tau) \simeq \frac{\chi_{2}(k)}{\sqrt{2 a_{\rm i}^2 \mu \tau^3}} \cos{\biggl(
\frac{\mu \tau^2}{2} + \gamma\biggr)}.
\label{solchi2}
\end{eqnarray}
The matching will allow a determination of the precise amplitudes and
phases of $\sigma$ (and $\chi_{k}$) in terms of $\sigma_{\rm i}$ (and
$\chi_{\rm i}(k)$).

\begin{figure}[t]
\centerline{\epsfxsize = 12 cm  \epsffile{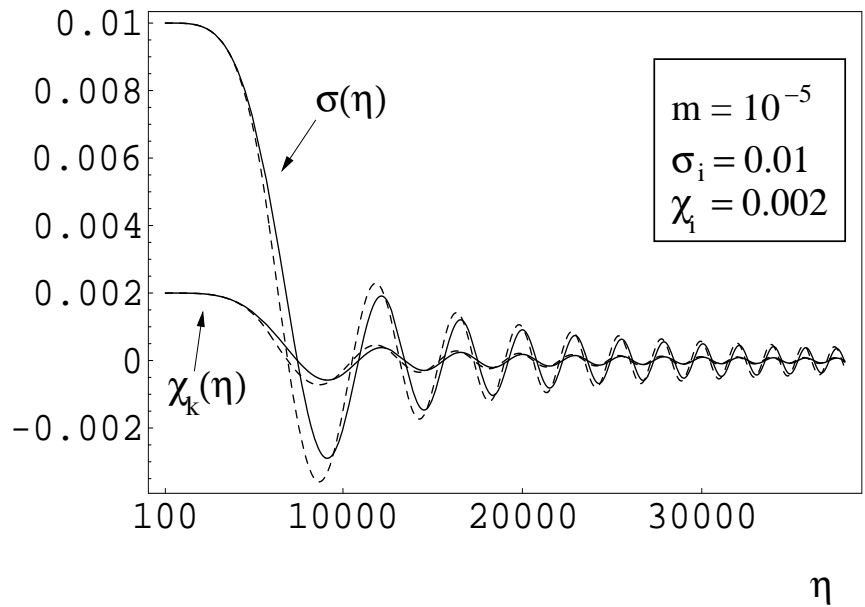}}
\vskip 3 mm
\caption[a]{ The exact, numerical evolution of  $\sigma$ and $\chi_{k}$
(full curves) is compared with the interpolating solution (dashed curves)
obtained by matching  the slow-roll solutions (\ref{sigsr}) and (\ref{chsr})
with  the WKB approximated solutions (\ref{solsig2}) and (\ref{solchi2}),
valid  during the radiation-dominated oscillations of the axion.}
\label{SR5}
\end{figure}

As an application of this technique let us consider the 
example of the quadratic potential,  using the slow-roll solutions
for $\tau < \tau_{\rm m}$. 
The result of this exercise is reported in Fig. \ref{SR5} where, with the
full curves, we illustrate the numerical  results (coinciding exactly
 with the analytical solutions). With the dashed curves we show the
interpolating solutions  obtained by matching Eqs.
(\ref{sigsr}) and (\ref{chsr}) (obtained  in the slow-roll approximation) 
with the WKB solutions (\ref{solsig2}) and (\ref{solchi2}),  valid in the
oscillating regime.  

\begin{figure}
\centerline{\epsfxsize = 12 cm  \epsffile{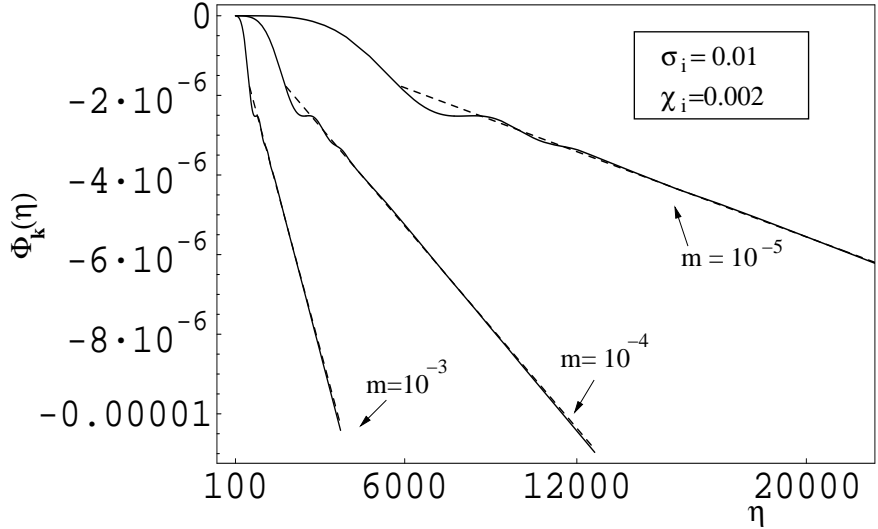}}
\vskip 3 mm
\caption[a]{The results of the  numerical integrations for the Bardeen 
potential are illustrated by the full curves for a quadratic potential and
for different values of the mass. The dashed lines represent the linear
fit of Eq. (\ref{regr}).}
\label{SR6a}
\end{figure}

The time evolution of $\sigma(\eta)$ and 
$\chi_{k}(\eta)$  explains why,
for $\tau \geq \tau_{\rm m}$ (i.e. after the slow-roll regime where
$\Phi_k \sim a^4$ according to Eq. (\ref{phgen})), the Bardeen potential
enters  a  phase of linear evolution (in conformal time). 
This feature is illustrated in Fig. \ref{SR6a},  
where we report the numerical results for the  evolution of the
Bardeen potential, computed for different values of the axion mass.

An analytical estimate of  
the slope of the linear regression for $\Phi_{k}$, after 
the end of the radiation-dominated slow roll,
can be obtained from the Hamiltonian 
constraint  (\ref{hamp}),
which can be recast in the following form: 
\begin{equation}
\frac{ \partial}{\partial \tau} ( \tau^3 \Phi_{k}) = - \frac{\tau^4}{12}
\biggl[ 
\frac{\partial\sigma}{\partial\tau}  \frac{\partial\chi_{k}}{\partial\tau}
+ \mu^2 \tau^2 \sigma  \chi_{k} \biggr],
\label{hamred}
\end{equation} 
where we only assumed a quadratic form for the axion potential. By
using the WKB solutions  (\ref{solsig2}) and (\ref{solchi2})
we obtain
\begin{equation}
\Phi_{k} = - \frac{ \sigma_2 \chi_{2}(k)}{96 a_{\rm i}^2} ~\mu \tau - 
\frac{ \sigma_2 \chi_2(k)}{384 a_{\rm i}^2\mu\tau^3}
\biggl[  - 12 \cos{( 2 \gamma + \mu\tau^2)} + 18 \int^{\mu\tau^2/2}  
\frac{dx}{x}  \cos{(x + \gamma)}\biggr],
\label{integr}
\end{equation}
where the integral can be expressed in terms of ${\rm Ci}(w) =
-\int_{w}^{\infty}  \frac{\cos{x}}{x} dx$ and  
${\rm Si}(w) = \int_{0}^{w} \frac{\sin{x}}{x} dx$, and we have assumed 
$\beta=\gamma$. 
The oscillating terms are suppressed by $\tau^{-3}$  and can be
neglected  (in agreement with the numerical results of Fig. \ref{SR6a}), 
since we are considering the regime $\tau > \tau_{\rm m} \gg 1$. On top
of the oscillating terms, the amplitude of the term responsible
for the linear growth can be extracted from the numerical solutions by
fitting their asymptotic behaviour with the line
\begin{equation}
\Phi_{k}(\eta) = \sigma_{\rm i} \chi_{\rm i} [ - \epsilon_2 - \epsilon_3 
\sqrt{\mu} \tau],~~~~~~ \tau > \tau_{\rm m}.
\label{regr}
\end{equation}
In the case of a quadratic
potential we obtain 
\begin{equation}
\epsilon_2 \simeq 0.001, ~~~~~~~~~~~\epsilon_3 \simeq 0.0437.
\label{est}
\end{equation}

The accuracy of this result can be appreciated from Fig. \ref{SR6a}, 
where the dashed lines (barely distinguishable from the  numerical
solutions)  are plotted according to Eqs. (\ref{regr}), (\ref{est}).
A different  form of the potential will not affect the angular coefficient
of  the regression (which is  determined by the phase of the
radiation-dominated  oscillations), but only the constant $\epsilon_2$. 

It may be interesting to look also at the analytical estimate of
$\epsilon_{3}$, for a quadratic potential. 
 The initial amplitudes
$\sigma_2$ and $\chi_{2}(k)$ of Eq. (\ref{integr}) are determined
by the  large argument limit of the exact solutions,  Eqs.
(\ref{wkbsig}) and  (\ref{chlarge}). In this case we get 
\begin{equation}
\epsilon_{3}^{\rm th} = \frac{\Gamma^2(5/4)}{6 \pi} = 0.0435,
\label{est2}
\end{equation}
in excellent agreement with (\ref{est}). For a generic potential, 
we could also determine $\epsilon_{3}$ by adopting an approximate
procedure, i.e. by taking the slow-roll solutions  for $\chi_{k}$ and
$\sigma$ from Eqs. (\ref{sre}) and (\ref{chev}), and matching them 
in $\tau_{\rm m}$ to the  WKB solutions
(\ref{solsig2}) and (\ref{solchi2}), in order to determine amplitude and
phase. 

The linear growth of the Bardeen potential continues  until  
$\rho_{\sigma}$, decreasing as $a^{-3}$, equals $\rho_{\rm r}$. This
happens  at a time $\tau_\sg$ such
that  
\begin{equation}
12 {\cal H}^2 \simeq a^2 m^2 \sigma^2.
\label{eq2}
\end{equation}
The expansion of Eq. (\ref{wkbsig}) for $\mu \ll 1$  then gives
\begin{equation} 
\tau_{\sigma} = \frac{\epsilon_4}{\sigma_{\rm i}^2~ \sqrt{\mu}}, 
~~~~~~~~~~~~
\epsilon_4 = \frac{3 \pi }{2 \Gamma^2(5/4)}\simeq 5.74. 
\label{tsig2}
\end{equation}
From Eq. (\ref{regr}) we can then finally obtain the value of the Bardeen
potential, at the onset  of the phase of $\sigma$-dominated
oscillations. By using the value of $\tau_\sg$ given by Eq. (\ref{tsig2}), 
the result is 
\begin{equation}
 \Phi_{k}(\eta_{\sigma}) \simeq \epsilon_{5} 
\frac{ \chi_{\rm i}(k)}{\sigma_{\rm i}},
~~~~~ \epsilon_{5} = \epsilon_4 \epsilon_3 \simeq  0.25.
\label{finph}
\end{equation}

\subsection{The axion-dominated oscillations}

Using standard techniques suitable for the 
oscillating regime \cite{tur,mfb}, 
Eqs. (\ref{ham}) and (\ref{kg}) can be solved and, 
in the case of a  quadratic potential, the oscillating 
terms lead 
to a geometry that reproduces (but only on the average) a 
matter-dominated Universe. The oscillating corrections 
will be suppressed for large (cosmic or conformal) times, and   can be
easily computed in the cosmic-time gauge, where: 
\begin{eqnarray}
&&\dot{H} = - \frac{1}{4} \dot{\sigma}^2,
\nonumber\\
&& H^2 = \frac{1}{12} \biggl[ \dot{\sigma}^2 + m^2 \sigma^2 \biggr],
\nonumber\\
&& \ddot{\sigma} + 3 H \dot{\sigma} + m^2 \sigma =0.
\end{eqnarray}
Using the auxiliary variable $Z= d \sg/d\ln{a}$, which satisfies
\begin{equation}
\frac{d Z}{d\sigma} = -
3 \bigg( 1 - \frac{Z^2}{12}\biggr)\biggl( 1 + 4 \frac{Z}{\sigma}\biggr), 
\label{red}
\end{equation}
and defining two angular variables $(r, \theta)$,
\begin{equation}
Z = \sqrt{12} \sin{\theta} ,~~~~\sigma= r\cos{\theta}, 
\end{equation}
(such that $H=mr/\sqrt{12}$), 
the following two equations are obtained: 
\beq
 \dot{\theta} = - \frac{\sqrt{3}}{4} m r \sin{2\theta} - m,
~~~~~~~~~~~
\dot{r} = -\frac{\sqrt{3}}{2} m r^2 \sin^2{\theta}. 
\eeq
They can be solved,  and the solutions expanded for large times at any 
order in $1/t$. 

A similar procedure can be carried out in  conformal time. 
Equations (\ref{kg})--(\ref{ham}) are equivalent to the following set of
equations  
\beq
 r' = - \frac{\sqrt{3}}{2} m a_{\rm f} \eta^2 r^2 
\sin^2{\biggl(\frac{m a_f \eta^3}{3}\biggr)},
~~~~~~~~~
 \sigma = r \cos{\biggl(\frac{m a_f \eta^3}{3}\biggr)}
\eeq
(where $a_f$ is an appropriate dimensionful integration constant), and
their solution leads to the expansion 
\begin{eqnarray}
&& a(\eta) \simeq a_{\rm f} \biggl[
\eta^2   - \frac{3}{2 a_{\rm f}^2 m^2 \eta^4} 
\cos{\biggl(\frac{2~m a_{\rm f} \eta^3}{3}\biggr)} + 
{\cal O}(\frac{1}{\eta^5})\biggr],
\nonumber\\
&& {\cal H}(\eta) \simeq
\biggl[ \frac{2}{\eta} + \frac{3}{ m a_{\rm f} \eta^{4}}
\sin{\biggl(\frac{2~m a_{\rm f} \eta^3}{3}\biggr)} 
+ {\cal O}(\frac{1}{\eta^5}) \biggr],
\nonumber\\
&& \sigma(\eta) \simeq \frac{4 \sqrt{3}}{m a_{\rm f} \eta^3} 
 \cos{\biggl(\frac{m a_{\rm f} \eta^3}{3}\biggr)} + {\cal O}(\frac{1}{\eta^5}).
\label{oscback}
\end{eqnarray}
A posteriori, as a cross-check,  
Eqs. (\ref{oscback}) can be inserted into
 Eqs. (\ref{dyn})--(\ref{ham}), to see that all  terms 
up to ${\cal O}(1/\eta^5)$ cancel, as expected. 

In the phase dominated by the oscillating axion the effective
gravitational source is pressureless, on the average. By inserting the
condition $\langle p_{\sigma}\rangle =0 $ into the background and
perturbation equations, we get 
\begin{equation}
\langle \delta_{\sigma}(k) \rangle \sim - 2 \langle \Phi_{k}\rangle,
~~~~~~~~~~~
\langle \delta_{\rm r}(k) 
\rangle \sim - 2 \langle \delta_{\sigma}(k) \rangle,
\label{relph}
\end{equation}
which can be inserted into Eq. (\ref{zetaex}), 
obtaining
\begin{equation}
\langle \zeta_{k} \rangle 
\simeq \frac{5}{6} \langle \delta_{\sigma}(k) \rangle \simeq - 
\frac{5}{3} \langle 
\Phi_{k}\rangle.
\end{equation}
when $\langle p_{\sigma}\rangle \sim 0$ and $\rho_{\sigma} 
\gg \rho_{\rm r} $.

Since $p_{\sigma} $ vanishes only on the  average,  more accurate 
solutions have to be supplemented  by oscillating corrections.
Using Eqs. (\ref{oscback}) an approximate
form of the perturbations in the oscillating 
regime can be obtained. 
One finds that $\delta_{\sigma}(k)$ and $\Phi_k$ are  almost constant
(up to oscillations), i.e.
\begin{eqnarray}
&&\delta_{\sigma}(\eta) \simeq -\frac{1}{2} \Phi_0(k)\biggl[1 
-  \cos{\biggl(\frac{2 a_{\rm f} m \eta^3}{3}\biggr)} \biggr], 
\label{dsosc}\\
&&\Phi_{k}(\eta) \simeq \Phi_0(k)\biggl[ 1  - 
\frac{1}{  m a_{\rm f} \eta^3} \sin{\biggl(\frac{2}{3} 
m a_{\rm f} \eta^3\biggr)} -
\frac{4 }{(m a_{\rm f} \eta^3)^2 } \cos{ 
\biggl(2\frac{m a_f \eta^3}{3}\biggr)}  \biggr],
\label{phiosc}\\
&& \chi_{k}(\eta) 
\simeq \chi_{0}(k)\biggl[
 \sin{\biggl(\frac{m a_{\rm f} \eta^3}{3}\biggr)} + 
\frac{ 3 }{ m a_{\rm f} \eta^3}  
\cos{\biggl(\frac{m a_{\rm f} \eta^3}{3}\biggr)}\biggr],
\label{chosc}
\end{eqnarray}
where 
\begin{equation}
\Phi_0(k) = \langle \Phi_{k}(\eta) \rangle,~~~~~~~~~\eta_\sg<\eta <
\eta_{\rm d}, 
\end{equation} 
and
\begin{equation}
\chi_{0}(k) = - \frac{4}{\sqrt{3}} \Phi_{0}(k).
\end{equation}
The above solutions 
satisfy the evolution equations of the fluctuations up to 
${\cal O}(\eta^{-5})$.

\subsection{The axion decay and the subsequent radiation-dominated
phase}

When the decay rate of the axion equals the cosmological expansion
rate, energy  is  transferred from the coherent oscillations of $\sigma$
to the radiation produced by the axion decay. The radiation produced thanks 
to the decay of the axion will quickly dominate the expansion and the second 
radiation-dominated phase will take place.

The Bardeen potential prior to decay  is given by Eq. 
(\ref{phiosc}) while,  after the decay, its evolution equation 
(\ref{phred}) reduces to 
\begin{equation}
\Phi'' + 4 {\cal H} \Phi' + 2 ( {\cal H}^2 + {\cal H}') \Phi - 
\frac{1}{3} \nabla^2 \Phi = 0, ~~~~~~ \eta >\eta_{\rm d},
\label{radbard}
\end{equation}
and the corresponding exact solution can  be expressed
as \cite{mfb}  
\begin{equation}
\Phi_{k}(\eta) = \frac{1}{\eta^3} \biggl[ B_{1}(k) ( \omega\eta
 \cos{\omega\eta}  - \sin{\omega\eta} ) + B_{2}(k)
 ( \omega\eta \sin{\omega\eta} + \cos{\omega\eta} )\biggr],
~~~ \eta>\eta_{\rm d},
\label{f1}
\end{equation}
where $\omega = k/\sqrt{3}$. In the sudden approximation, the  
two (dimensional) arbitrary constants $B_1(k)$ and $B_2(k)$ can be 
uniquely fixed by matching, at the decay time $\eta_{\rm d}$,  the solutions 
(\ref{phiosc}) and (\ref{f1}) together  with their first derivatives.
 The terms containing 
$(k\eta_{\rm d})$  are  small and negligible for
modes that are outside the horizon  at the time of the axion decay,
i.e. for the  ones relevant to the  physics of the observed CMBR
anisotropies. Furthermore, terms proportional to inverse powers of 
$m/H_{\rm d}\sim M_{\rm P}^2/m^2$ are 
also small and can be consistently neglected.
 Hence, up to subleading terms,  the  final value of
the Bardeen potential can be written as  
\begin{equation} 
\Phi_{k}(\eta)
= \Phi_{0}(k)  \left[  2\cos{\left(\frac{2 \b}{3}\right)} - 3\right]   \biggl[
\frac{\cos{\omega \eta}}{(\omega \eta)^2} - 
\frac{\sin{\omega\eta}}{(\omega\eta)^3} \biggr],  
\label{bardint2}
\end{equation}
where  $\b = m \eta_{\rm d} a(\eta_{\rm d})
\sim m/H_{\rm d} \sim M_{\rm P}^2/m^2$.
The  $\b$-dependent prefactor is a consequence of the 
approximation of sudden decay where the axion field 
is assumed to decay at a specific time $\eta_{\rm d}$. This 
sudden approximation also neglects the possible (exponential) damping 
of the oscillations in $\Phi_{k}$ arising in  
 Eq. (\ref{phiosc}).

It will now be shown 
that  the $\beta$-dependent prefactor is an artefact of the sudden 
approximation.
In a realistic model of decay, in fact, the 
energy-momentum tensors of the radiation fluid and of the 
axion will not be separately conserved, because of their relative
coupling induced by the friction term $\Ga (\sg'/a)^2$, which leads, in
cosmic time, to the generalized conservation equations:
\begin{eqnarray} 
&&\dot{\rho}_{\sigma} + (3 H + \Gamma)
(\rho_{\sigma} + p_{\sigma})=0, \nonumber\\ && \dot{\rho}_{\rm r} + 4
H \rho_{\rm r} - \Gamma(\rho_{\sigma} + p_{\sigma})  =0.
\label{dec2}
\end{eqnarray}
The fluctuations $\chi_{k}$ will experience 
a similar damping,
\begin{equation}
\chi'' + (2 {\cal H}+\Gamma a) \chi' - \nabla^2 \chi + 
\frac{\partial^2 V}{\partial\sigma^2} a^2 \chi - 4 \sigma' \Phi' + 2 
\frac{\partial V}{\partial \sigma }a^2 \Phi =0,
\label{chiGamma}
\end{equation} 
while the $\Phi_{k}$ evolution will still be described by Eq. (\ref{phred}).
This treatment of the damping of the fluctuations 
was suggested in \cite{us} (see also \cite{mwu}).
The effect of $\Ga$ is, primarily, to induce a 
damping in  the oscillations 
of the background and of the axion fluctuations according to Eqs. 
(\ref{dec2})  and (\ref{chiGamma}).
Moreover, the (damped) fluctuations of the axionic field will also 
influence the dynamics of $\Phi_{k}$ according to Eq. (\ref{phred}). 
The time-dependent oscillations of Eq. (\ref{phiosc})  (occurring 
in the absence 
of friction) will then be further suppressed if $\Gamma \neq 0$ 
(more details
will be given in the following section).  
As a consequence,  the $\b$-dependent correction tends to 
disappear from  Eq. (\ref{bardint2}), leading to the final result 
\begin{equation}
\Phi_{k}(\eta) = 3 \Phi_{0}(k)  
\biggl[ 
\frac{\sin{\omega\eta}}{(\omega\eta)^3}
- \frac{\cos{\omega \eta}}{(\omega \eta)^2} \biggr]. 
\label{bardint3}
\end{equation}
In the equation for $\Phi_{k}$ 
 the effect of the finite duration $\Gamma^{-1}$ 
is then equivalent to averaging 
over the decay time. 
At the end of the following section, numerical examples of  the decay 
will be discussed in detail.

\renewcommand{\theequation}{5.\arabic{equation}}
\setcounter{equation}{0}
\section{Background and perturbation equations for
$\sigma_{\i} > 1$}

\begin{figure}
\centerline{\epsfxsize = 12 cm  \epsffile{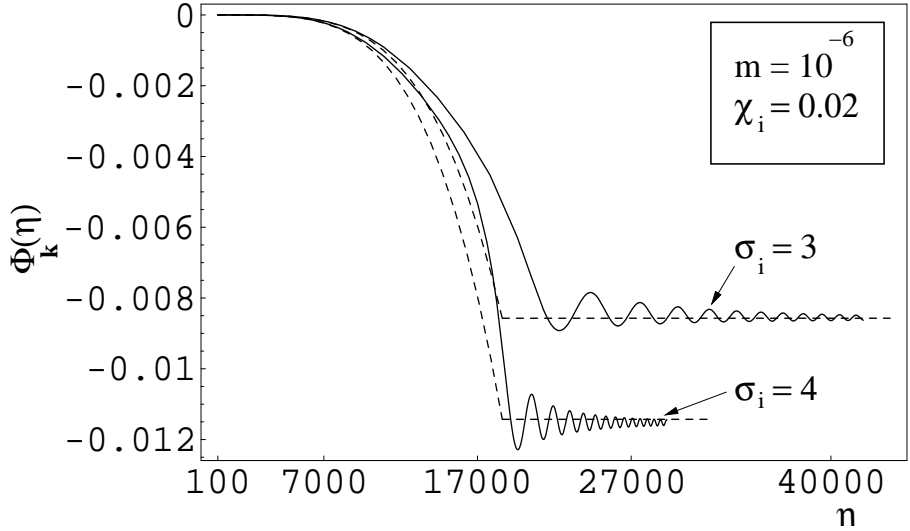}}
\vskip 3 mm
\caption[a]{The full  curves are the results of numerical integrations
for the case $\sigma_{\rm i} >1$. 
The dashed curves correspond to the approximated results of 
Eqs.  (\ref{phsr}) and (\ref{57}).}
\label{SR3a}
\end{figure}

If $\sigma_{\rm i}>1$, the epoch
of axion domination precedes the oscillation epoch. The previous
solutions for the slow-roll regime are still valid, and the axion starts
dominating when
 \begin{equation} 
\rho_{\rm r} a^2 = 6 {\cal H}^2
\simeq V a^2, 
\label{dom}
\end{equation}
i.e., for a quadratic potential, when
\begin{equation}
\tau=\tau_{\sigma} \simeq \frac{(12)^{1/4}}{\sqrt{\mu\sigma_{\rm i}}}.
\label{tsig}
\end{equation}
If $\sigma_{\rm i} \simeq 1$, then
\begin{equation}
\tau_{\rm m} \simeq \tau_{\sigma}
,\label{hier1}
\end{equation}
the axion oscillates almost immediately after becoming dominant,  and
the amplitude of the Bardeen potential
at the onset of the oscillatory phase
 is  obtained from Eq. (\ref{phsr}) as
\begin{equation}
\Phi_{k} ( \eta_{\rm m}) \simeq - \frac{\mu^2}{84} \chi_{\rm i}(k) 
\sigma_{\rm i}  \tau_{\rm m}^4 = - {1\over 7} {\chi_{\rm i}(k) \over
\sigma_{\rm i}} \simeq  - {1\over 7} \chi_{\rm i}(k) . 
\label{mean1}
\end{equation}
If the initial value $\sg_{\rm i}$ is larger than $1$, but not too large, then 
Eqs. (\ref{hier1}) and (\ref{phsr})  are still valid, but Eq. (\ref{mean1}) is
to be multiplied by the factor $\sg_{\rm i}^2$, arising from a short period of
axion dominance toward the end of the slow-roll evolution (see
below). This effect, for moderate values of $\sg_{\rm i}$, is illustrated in 
Fig \ref{SR3a} where, for the given parameters of the plot, the
final amplitude of the Bardeen potential is estimated as 
\begin{equation}
\Phi_{k} ( \eta_{\rm m}) \simeq - \epsilon_{6} \chi_{\rm i}(k)
\sigma_{\rm i}, ~~~~~~~~~~~
\epsilon_6 = 0.143
\label{57}
\end{equation}
still in good agreement with the approximate value $1/7$ of Eq. 
(\ref{mean1}). 

If $\sigma_{\rm i} \gg 1$, then $\tau_{\rm m} \gg \tau_{\sigma}$,
and a phase of inflationary expansion dominated by the axion potential
will take place between $\tau_{\sg}$ and 
$\tau_{\rm m}$. During this phase the axion slowly rolls, the radiation
energy-density is quickly diluted  as $\rho_{\rm r} \sim a^{-4}$, and
the time evolution of the fluctuations is correspondingly 
modified. The Hamiltonian and momentum constraints 
(\ref{hamp})  and (\ref{momp}) can now be combined to give  
\begin{equation}
4\nabla^2 \Phi_{k} = 3 {\cal H} \sigma' \chi_{k} - \Phi_{k} 
{\sigma'}^2 + \sigma' \chi_{k}' + 
V_{,\sigma} a^2 \chi_{k}, 
\label{comb}
\end{equation}
and the speed of sound of Eq. (\ref{sps}) becomes
\begin{equation}
c_s^2 \simeq 1 + \frac{2 a^2}{3 {\cal H} \sigma'} V_{,\sigma}, 
\label{sps2}
\end{equation}
from which, using Eq. (\ref{dpnadex}), 
\begin{equation}
\delta p_{\rm nad} = \frac{2 V_{\sigma}}{3 {\cal H} \sigma'} 
\biggl[ \Phi_{k} 
{\sigma'}^2 + \sigma' \chi_{k}' - 3 {\cal H} \sigma' \chi_{k} - a^2 
V_{,\sigma} 
\chi_{k}
\biggr].
\label{dpnadinf}
\end{equation}
The combination of Eqs. (\ref{comb}) and (\ref{dpnadinf}) leads to
\begin{equation}
\delta p_{\rm nad} = \frac{ 8 V_{,\sigma}}{3 {\cal H} \sigma'} \nabla^2\Phi_{k}.
\end{equation} 
Hence, from Eq. (\ref{zetapr}), we get that $\zeta_{k} ' \sim 0$ at large
scales.  

According to its definition, on the other hand, the constancy of
$\zeta_{k}$  implies, in cosmic time, that 
\begin{equation}
\Phi_{k} \biggl( \frac{ 2 + \alpha_1}{ 1 + \alpha_1}\biggr) + 
\frac{\dot{\Phi}_{k}}{H} 
\frac{1}{1 + \alpha_1} 
\label{constant}
\end{equation}
is also a constant, where 
\begin{equation}
\alpha_{1} = - {\dot{H}}/{H^2} .
\end{equation}
It follows that, during inflation, we can parametrize the evolution of
$\Phi_k$, to lowest order, as  
\begin{equation}
\Phi_{k} = A \biggl({\dot{H}}/{H^2}\biggr),
\label{Phinf}
\end{equation}
where $A$ is a constant controlled by the value of the Bardeen 
potential at the beginning  of inflation. Assuming a quadratic potential
for $\tau < \tau_{\sigma}$  we have, from Eq. (\ref{mean1}),  $A
 \simeq -(1/7)
(\chi_{\rm i}(k)/\sigma_{\rm i})(H^2/\dot H)_{\tau_\sg}$. By using the
dynamics of slow-roll inflation, 
\begin{eqnarray}
&& H^2 \simeq \frac{m^2}{12} \sigma^2,
\label{srin1}\\
&& \dot{\sigma} \simeq - \frac{m^2}{3 H} \sigma, 
\label{srin2}
\end{eqnarray}
we can deduce that 
\begin{equation}
\frac{\dot{H}}{H^2} \simeq -\frac{4}{\sigma^2}.
\label{hdoh}
\end{equation}
Using Eq. (\ref{hdoh}) 
we finally obtain the Bardeen potential at the onset of the
phase of  $\sigma$-dominated oscillations:  
\begin{equation}
\Phi_{k}(\tau_{\rm m}) \sim \sigma_{\rm i} \chi_{\rm i}(k),
\label{515}
\end{equation} 
where we used the expression for $A$ given above and the fact that 
$\sigma(\tau_{\sigma}) = \sigma_{\rm i}$ 
and $\sigma(\tau_{\rm m}) \sim {\cal O}(1)$.
For $\tau>\tau_{\rm m}$ the axion eventually oscillates, and the subsequent
evolution has the same features as already discussed in the previous
section, for  the case $\sigma_{\rm i}<1$. 

\begin{figure}
\centerline{\epsfxsize = 12 cm  \epsffile{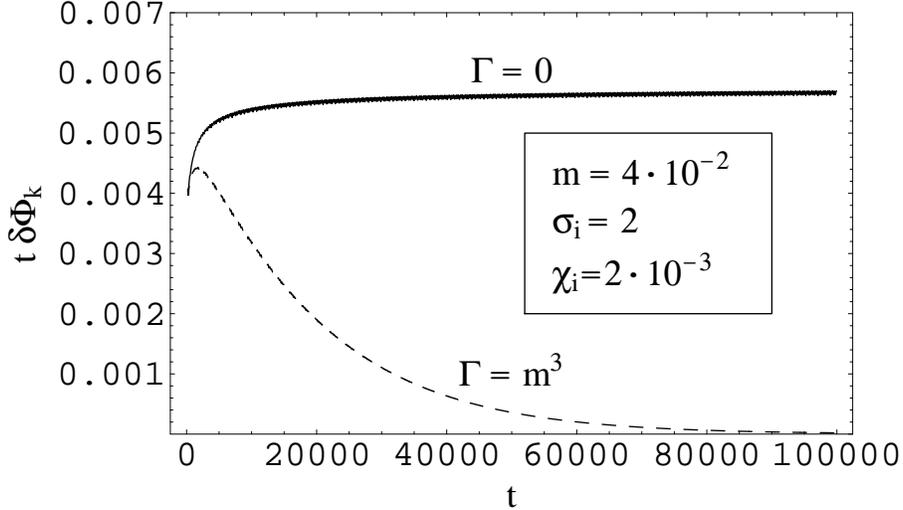}}
\vskip 3 mm
\caption[a]{Time evolution of the amplitude of the $\delta\Phi_k$ 
oscillations  (multiplied by $t$ in cosmic time), with and without the
damping term due to the axion decay.}  
\label{Fdec1}
\end{figure}

\begin{figure}
\centerline{\epsfxsize = 12 cm  \epsffile{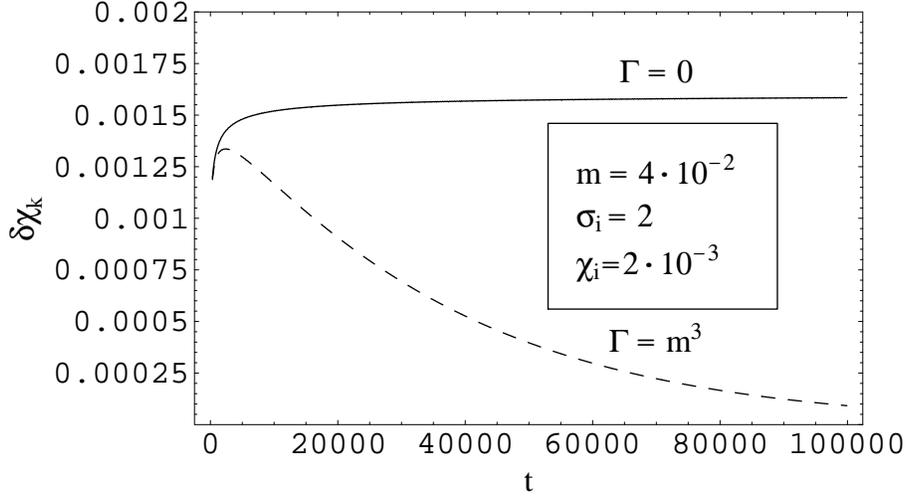}}
\vskip 3 mm
\caption[a]{Time evolution of the amplitude of
the $\delta\chi_{k}$ oscillations, with and without the
damping term due to the axion decay.} 
\label{Fdec2}
\end{figure}

In the following, the dynamics of the decay will be investigated 
numerically, and Eq. (\ref{phred}) will be solved together 
with Eqs. (\ref{dec2}) and (\ref{chiGamma}).
In order to illustrate the  results let us recall 
that, in the absence of friction ($\Gamma =0$ in Eqs. (\ref{chiGamma})), 
the evolution of $\chi_{k}$ and $\Phi_{k}$, during the 
axion-dominated  oscillations,  is given by Eqs. (\ref{phiosc}) and
(\ref{chosc}). In particular,   
\begin{equation}
\Phi_{k}(\eta) \simeq \Phi_{0}(k) + \delta \Phi_{k}(\eta),
\label{phdecdef}
\end{equation}
where $\delta\Phi_{k}(\eta)$ is an oscillating  function\footnote{In
order  to avoid confusion we note that  $\delta\Phi_{k}$ and, in the
following, $\delta\chi_{k}$,  are not the power spectra of $\Phi$ and
$\chi$.}  decaying as  $\eta^{-3} \sim t^{-1}$. The frequency of
oscillation  of $\delta\Phi_{k}$ is controlled by the axion mass. In
analogy with Eq. (\ref{phdecdef}) we can also define $\delta \chi_{k}$ 
which, for $\Gamma =0$, corresponds to the oscillating function
appearing in Eq. (\ref{chosc}). The evolution of $t\delta\Phi_{k}$ and of 
$\delta\chi_{k}$, 
 for $\Gamma=0$, is represented by the full bold curves of 
 Figs. \ref{Fdec1} and \ref{Fdec2}. Notice that $\delta\Phi_{k}$ and
$\delta\chi_{k}$ oscillate  very fast and that, for our illustrative
purpose, we have plotted  their amplitudes calculated as the 
average of the semi-difference
between the maximum and the minimum of each oscillation, and the
semi-difference between the successive maximum and the same
minimum.

If the oscillations 
of $\delta\Phi_{k}$ are only suppressed by a power-law
function of time, we have seen
that there are mass-dependent terms that appear in the amplitude of
the  Bardeen potential after the decay. 
The  integration of Eqs. (\ref{phred}) and of
(\ref{dec2}) and (\ref{chiGamma}) shows however that, with the inclusion
of the  appropriate friction terms (due to the decay) into the
energy-momentum conservation equations, 
the oscillations in $\delta\Phi_{k}$ and $\delta\chi_{k}$ are
exponentially suppressed and, in such a case, no mass-dependent
correction is left in the amplitude of the Bardeen potential (the
asymptotic, constant value of $\Phi_{k}$ is however unaffected by
such a damping mechanism). The damping of the oscillations, on a time
scale of order $\Gamma^{-1}$, is illustrated by the dashed curves of
Figs.  \ref{Fdec1} and \ref{Fdec2}.  

If we  compare the sudden-decay approximation, discussed 
in the previous section,  
 with the numerical results of Figs. \ref{Fdec1} and \ref{Fdec2}, we 
see that the finite duration of the decay process can be physically
represented as a dynamical average to zero of the oscillatory terms in
the evolution of $\Phi_k$. In view of these results, when matching
$\Phi_k$ to the post-decay phase, we should take into account the fact 
that all
the derivatives of $\Phi_k$ are exponentially suppressed with respect
to $\Phi_k/t_{\rm d}$,  and thus can be safely neglected. This 
leads to the result reported in  Eq. (\ref{bardint3}). 

\renewcommand{\theequation}{6.\arabic{equation}}
\setcounter{equation}{0}
\section{Large-scale adiabatic fluctuations}

In order to discuss the direct impact of our results on the possible
generation of the observed  CMBR anisotropies, the
evolution  of the large-scale metric fluctuations should be followed
down to the matter-dominated phase, for all times  $\eta>\eta_{\rm
eq}$.  In particular, the phase and the amplitude  of the Bardeen
potential prior to $\eta_{\rm eq}$ will fix the initial conditions for the
subsequent evolution of  the inhomogeneities, and will be crucial to
determine whether they are of adiabatic or isocurvature nature.

We recall that, after the axion decay, the amplitude of the Bardeen
potential has been computed as 
\begin{equation}
\Phi_{k}(\eta) = 3 \Phi_{0}(k)  
\biggl[
\frac{\sin{\omega\eta}}{(\omega\eta)^3}- 
\frac{\cos{\omega \eta}}{(\omega \eta)^2} \biggr],
~~~~~\eta \leq \eta_{\rm eq},
\label{bardint4}
\end{equation}
where, as in the previous section, $\omega= k/\sqrt{3}$.
For $\eta > \eta_{\rm eq}$, matter domination sets in, the background
satisfies $ 2 {\cal H}' + {\cal H}^2=0$, so that the evolution of the
Bardeen potential (outside the horizon) is described by 
\begin{equation} 
\Phi_{k}'' + 3 {\cal H}\Phi_{k}' =0,~~~~~{\cal H} =
\frac{2}{\eta}, \label{mdeq}
\end{equation}
whose solution can be written as 
\begin{equation}
\Phi_{k}(\eta) = A(k) + \frac{B(k)}{\eta^{5}},~~~~\eta\geq \eta_{\rm eq}.
\label{mdsol}
\end{equation}
Imposing the continuity of the solutions (\ref{bardint4}) and (\ref{mdsol}) 
and of their first derivatives at $\eta= \eta_{\rm eq}$ one obtains:
\begin{eqnarray}
&& A(k) = \frac{3 \Phi_{0}(k)}{5 x_{\rm eq}^3}[ - 2 x_{\rm eq} 
\cos{x_{\rm eq}} + 
(x_{\rm eq}^2 + 2) \sin{x_{\rm eq}}], 
\label{Ak}\\
&&B(k) = \frac{3 \Phi_{0}(k) \eta_{\rm eq}^5}{5  x_{\rm eq}^3}
[ 3 x_{\rm eq} \cos{x_{\rm eq} } + 
(x_{\rm eq}^2 - 3) \sin{ x_{\rm eq}}],
\label{Bk}
\end{eqnarray}
where $x_{\rm eq} = \omega \eta_{\rm eq} \equiv k 
\eta_{\rm eq} /\sqrt{3}$.
For scales that are outside the horizon prior to decoupling, 
 $x_{\rm eq} \ll 1$, and  Eq. (\ref{mdsol}) 
becomes
\begin{equation}
\Phi_{k}(\eta) = \Phi_{0}(k)\biggl[1 + \frac{ (k\eta_{\rm eq})^2 }{75}
\biggl( \frac{\eta_{\rm eq}}{\eta}\biggr)^{5}\biggr], 
~~~~~~~\eta> \eta_{\rm eq}.
\label{MD}
\end{equation}

For $\eta> \eta_{\rm eq}$ the decaying mode 
is highly suppressed, and we are then in the situation of constant
Bardeen potential  right after equality, with an amplitude $\Phi_{0}(k)$, 
which (recalling the previous results (\ref{finph}), (\ref{mean1}),
(\ref{515}))  is  completely determined by  the axion spectrum
and by  the initial conditions of the axion background. More precisely,
the final amplitude can be parametrized as follows
\begin{equation}
\Phi_{0}(k) \equiv \Phi_{k}(\eta_{\rm d}) \equiv - 
f(\sigma_{\rm i})  \chi_{\rm i}(k), 
\label{phif}
\end{equation}
where 
\begin{equation}
f(\sigma_{\rm i}) = c_1 \sigma_{\rm i} +\frac{c_2}{\sigma_{\rm i}} -
c_3,  
\label{fsig}
\end{equation}
and  
\beq
 c_1 \simeq 0.13,
~~~~~~~
 c_2 \simeq 0.25,
~~~~~~~
c_3 \simeq 0.01.
\label{c123}
\eeq
The above coefficients $c_{i}$  have been obtained
by  integrating numerically the evolution equations of the background 
and of the fluctuations for different values of $\sigma_{\rm i}$ (both 
larger and smaller than $1$). Then, following the 
hint of the analytical results obtained by solving the evolution piece-wise,
the final value of $\Phi_{k}(\eta)$ has been fitted with Eq. (\ref{fsig}), 
and the  values reported in Eq. (\ref{c123}) have been determined.
 
The value (\ref{fsig}) of the Bardeen potential provides the initial 
condition for the subsequent hydrodynamical evolution. 
Such evolution  will allow us to determine, in turn,  the precise value of
the temperature fluctuations through the Sachs--Wolfe effect.  In
particular, the modes that are outside the horizon for  $\eta_{\rm
eq}< \eta <\eta_{\rm dec} $  will determine the large-scale temperature
fluctuations relevant to the COBE observations. 

By perturbing the corresponding conservation equations on a  
matter-dominated background,  we obtain: 
\begin{eqnarray}
&& \delta_{\rm r}' - \frac{4}{3} \nabla^2 v_{\rm r} - 4 \Phi' =0,
\label{dr1}\\
&& v_{\rm r}' - \frac{1}{4} \delta_{\rm r} - \Phi = 0,
\label{ur1}\\
&& \delta_{\rm m}' - \nabla^2 v_{\rm m} - 3 \Phi' =0,
\label{dm1}\\
&& v_{\rm m}' + {\cal H} v_{\rm m} - \Phi =0,
\label{udm}
\end{eqnarray}
where $\delta_{\rm m}=\da \r_{\rm m}/\r_{\rm m}$ and $v_{\rm m}$, 
following the notation of the previous sections, are 
the gauge-invariant density contrast and velocity potential of the
matter fluctuations.  Also, in the above equations, 
\begin{equation} {\cal H} = \frac{2}{\eta},~~~~~\rho_{\rm m} a^2 =
\frac{24}{\eta^2}. 
\label{backdm}
\end{equation}
As already stressed at the beginning 
of this section, $\Phi$ is constant during the matter-dominated phase.
Using this property we can now work out the specific  relations
between the different fluid  variables, for 
modes that are outside the horizon  right after equality, so as to
explicitly check the adiabaticity of the fluid perturbations. 

The system of Eqs. (\ref{dr1})--(\ref{udm}) 
can be easily solved by going to Fourier space. For $v_{\rm m}$ we have
\begin{equation}
k v_{\rm m}(k) \simeq \frac{k\eta}{3} \Phi_0(k),~~~~~~~k\eta \ll 1.
\end{equation}
 Since $\vec{\nabla} v_{\rm m}$ (evaluated outside the horizon) contributes
directly to the Sachs--Wolfe effect, it is  important to notice 
that this term is subleading  with respect to the other contributions arising 
in the case of adiabatic fluctuations. 
We will indeed show that, unlike  $\vec{\nabla} v_{\rm m}$, which 
is suppressed, the contrast 
$\delta_{\rm r}$ is instead constant outside the horizon, and 
proportional to $\Phi_{0}(k)$.

Inserting 
$v_{\rm r}$ from Eq. (\ref{dr1}) into Eq. (\ref{ur1}) we get a 
decoupled equation for $\delta_{\rm r}$, namely, 
\begin{equation}
\delta_{\rm r}'' + \frac{k^2}{3} \delta_{\rm r} = - \frac{4}{3} k^2 
\Phi_{0}(k).
\label{decdr}
\end{equation}
The general solution  is
\begin{equation}
\delta_{\rm r}(k \eta) = A_1 \cos{\omega \eta} + B_1\sin{\omega\eta} + 
4 \Phi_{0}(k) [\cos{\omega\eta} - 1], 
\end{equation}
and the constants $A_{1}$ and $B_{1}$ 
can be determined by consistency with the other
equations and with the Hamiltonian constraint (\ref{hamp}) written  in the case 
of a matter-radiation fluid. The  final result is:
\begin{eqnarray}
&& \delta_{\rm r}(k,\eta) = \frac{4}{3}\Phi_{0}(k) \biggl[ \cos{\omega\eta} - 3\biggr]
\label{delrel}\\
&& k v_{\rm r}(k,\eta) = \frac{\Phi_0(k)}{\sqrt{3}} \sin{\omega\eta},
\label{urel}\\
&& \delta_{\rm m}(k,\eta) = - 2 \Phi_{0}(k)-\frac{\Phi_{0}(k)}{6} (k\eta)^2, 
\label{dmrel}\\
&& k v_{\rm m}(k,\eta) = \frac{ (k\eta)}{3} \Phi_{0}(k).
\label{uma}
\end{eqnarray}
Notice that, outside the horizon, $ k v_{\rm m} \equiv k v_{\rm r}$ as required by 
local thermodynamical equilibrium. Furthermore, for $k\eta \ll 1$, the velocities 
of the two fluids are proportional to $(k\eta)$.  When the modes are outside the horizon,
Eqs. (\ref{delrel}) and (\ref{dmrel}) imply that the density 
contrasts $\delta_{\rm r}$ and $\delta_{\rm m}$  are both 
constant and proportional according to 
\begin{equation}
\delta_{\rm r} \simeq (4/3) \delta_{\rm m}.
\label{dmrad}
\end{equation}
This result has a simple physical interpretation, and 
implies the adiabaticity of the fluid perturbations. 
The entropy per matter particle is indeed proportional to
 $S=T^3/n_{\rm m}$, where $n_{\rm m}$ is the number density of 
 matter particles and $T$ is the radiation temperature.
The associated entropy fluctuation, $\da S$, satisfies  
\begin{equation} 
\frac{\delta S}{S} = \frac{3}{4} \delta_{\rm r} - \delta_{\rm m},
\end{equation}
where we used the fact that $\rho_{\rm r} \sim T^4 $ and that 
$\rho_{\rm m} = m n_{\rm m}$, where $m$ is the typical mass 
of the particles in the matter fluid. Equation (\ref{dmrad}) thus implies  $\delta S/S
=0$, in agreement with the adiabaticity  of the fluctuations.

\subsection{Sachs-Wolfe effect and COBE scales}

The fluctuations of the Bardeen potential and of the 
radiation density contrast are sources of a slight temperature
difference between photons coming from different sky directions. This is the
essence of the Sachs--Wolfe effect \cite{sw}. In terms of the
gauge-invariant variables introduced in the present analysis, the
various contributions  to the Sachs--Wolfe effect, along the $\vec{n}$
direction, can be written as  \cite{mfb,dur} 
\begin{equation}
\frac{\Delta T}{T}(\vec{n},\eta_0,x_0) = \biggl[ \frac{\delta_{\rm r}}{4} 
+ \vec{n} \cdot \vec{\nabla} v_{\rm b} 
+ \Phi\biggr](\eta_{\rm dec}, \vec{x}(\eta_{\rm dec})) 
- \int_{\eta_0}^{\eta_{\rm dec}} (\Phi' + \Psi')(\eta , \vec{x}(\eta))
d\eta, 
\label{SW}
\end{equation}
where $\eta_0$ is the present time, and $\vec{x}(\eta)=\vec{x}_0-
\vec{n}(\eta-\eta_0)$ is the 
 unperturbed photon position 
at the time $\eta$ for an observer 
 in $\vec{x}_0$. The term  $\vec{v}_{\rm b}$ is the peculiar velocity 
of the baryonic matter component.
We are preliminarily interested in the effects of scales  still outside the
horizon at the time of the matter-radiation equality, which are the 
scales relevant to the observations of the COBE-DMR  experiment
\cite{smoo1,smoo2}. 
In order to correctly take into account the constraints 
imposed by the COBE normalization on the 
spectral amplitude of the Bardeen potential,  let us  compare the
relative weight of the different terms appearing in the  Sachs-Wolfe
formula (\ref{SW}).

From Eq. (\ref{urel}) we can see that, for our adiabatic initial 
conditions, the fluctuation 
in the  matter velocity potential is 
subleading for superhorizon scales, suppressed by the term $k\eta
\ll 1$ with respect to the constant values of   $\delta_{\rm
r}$ and $\Phi_{k}$.  Furthermore, since $\Phi'\simeq 0$ and $\Psi =
\Phi$,  the integrated  Sachs--Wolfe effect can also be neglected. By 
inserting Eq. (\ref{delrel}) into Eq. (\ref{SW}) we thus obtain the usual 
result for adiabatic fluctuations, namely
\begin{equation}
\frac{\Delta T}{T} (\vec{n},\eta_0,x_0) = \frac{1}{3}
\Phi (\eta_{\rm dec}, \vec{x}(\eta_{\rm dec})), 
\label{ff}
\end{equation} 
to be used  for the comparison of our theoretical predictions 
with the COBE normalization.

On the other hand, by taking the Legendre transform at the present 
time $\eta_0$, the  temperature  fluctuations of Eq. (\ref{SW}) can be
generally expanded into spherical harmonic functions, $Y_{\ell m}$, as 
\begin{equation} 
\frac{\Delta T}{T}(\vec{x}_0,\vec{n},\eta_0)
=\sum_{\ell,m} a_{\ell m}(\vec{x}_0) Y_{\ell m}(\vec{n}), 
\end{equation}
where the coefficients $a_{\ell m}$ define the angular power spectrum 
$C_\ell$ by 
\begin{equation}
\biggl\langle a_{\ell m}\cdot a_{\ell'm'}^*\biggr\rangle 
= \delta_{\ell\ell'}\delta_{mm'}C_\ell,
\label{anps}
\end{equation}
and determine the 
two-point correlation function of the temperature fluctuations,
namely
\begin{eqnarray}
\left\langle{\delta T\over T}(\vec{n}){\delta T\over T}(\vec{n}') \right\rangle_{{~}_{\!\!(\vec{n}\cdot
\vec{n}'=\cos\vartheta)}}&=&
\sum_{\ell \ell' m m'}\biggl\langle a_{\ell m}
 a_{\ell'm'}^*\biggr\rangle  Y_{\ell m}(\vec{n})  Y_{\ell'  m'}^*(\vec{n'})
\nonumber \\
& =& {1\over 4\pi}\sum_\ell(2\ell+1)C_\ell P_\ell(\cos\vartheta).
\end{eqnarray}
These coefficients $C_{\ell}$, in 
turn, are  related through Eq. (\ref{ff}), to the power spectrum of 
$\Phi_{0}(k)$,  and for $2 \leq \ell  \ll 100$ they can be expressed as
\cite{dur2} 
\begin{equation}
C_\ell \simeq 
 \frac{2}{9 \pi} \int_0^\infty \frac{dk}{k}
 \biggl\langle |\Phi_{0}(k)|^2
 \biggr\rangle k^3 j_\ell^2 [k(\eta_0-\eta_{\rm dec})].
\label{CL1}
\end{equation}

As already stressed, the spectrum of the Bardeen potential is fully
determined, in our context, by the initial spectrum of axionic
fluctuations amplified by  the pre-big bang
dynamics. A  self-contained derivation of such a spectrum, including
the mass contribution, is presented in Appendix A. 
Consider first the case of minimal  pre-big bang models, whose related
spectrum is reported  in Eq. (\ref{estimate}). The spectrum of curvature
perturbations  will then be, at large scales, 
\begin{equation}
k^3 \left| \Phi_0(k)\right|^2 = f^2(\sigma_{\rm i}) k^3 
\left|\chi_k\right|^2 =  f^2(\sigma_{\rm i}) \biggl(\frac{H_1}{M_{\rm P}}
\biggr)^2 
\left( k\over k_1\right)^{n-1}, ~~~~~~k<k_1,
\label{norm}
\end{equation}
where $k_1$ is the maximal amplified comoving frequency, i.e.,
in our conventions, the frequency at which only one 
axion per cell of phase-space is produced. 
From $k_1$, a typical curvature scale $H_1$  (which can be, at most, of
the  order of the string mass) can be obtained:
\begin{equation}
H_{1} = \frac{k_1}{a_{1}} \leq M_{\rm s}.
\label{H1}
\end{equation}
The particular value of the scale $H_1$ may be regarded 
as a phenomenological parameter of the chosen model of 
pre-big bang evolution. Even assuming, according to the standard lore
\cite{kap}, that $M_{\rm s}\sim 10^{-1}~M_{\rm P}$, still the exact
relation of $H_{1}$ to $M_{\rm s}$ depends on the detailed dynamics of a
highly  curved and strongly coupled background. The approach of the 
present investigation has been to include all the theoretical 
indetermination into $H_1$, trying to have a reasonable 
control of all the other numerical factors associated with 
the post-big bang evolution.
In Eq. (\ref{norm}) the particular value of $n$ depends upon the
specific model of pre-big bang evolution \cite{cew,mvdv}. In the case 
of a ten-dimensional model with an isotropic 
six-dimensional internal space, the line element 
can be written as 
\begin{equation}
ds^2 = dt^2 - a^2(t)\gamma_{i j} dx^{i} dx^{j} - 
b^2(t) \gamma_{a b} dy^{a} dy^{b},  
\end{equation}
where $i, j$ run over the three external space-like dimensions and 
$a, b$ run over the six internal dimensions. 
Defining as 
\begin{equation}
r = \frac{\dot{V}_{6} V_3}{2 V_{6}\dot{V}_{3}}
\end{equation}
the relative rate of variation of the external $V_{3} = a^3$ and internal 
$V_{6} = b^6$ volumes, the spectral index $n$ can be expressed as 
\cite{mvdv}  
\begin{equation}
n = \frac{ 4 + 6 r^2 - 2 \sqrt{ 3 + 6 r^2}}{ 1 + 3 r^2}.
\label{index}
\end{equation}
The case
of flat spectrum (i.e. $n=1$)
corresponds to the case $r=\pm 1$. If internal dimensions
are static (i.e. $r=0$), then $ n = 4  - 2 \sqrt{3}\simeq 0.53$.  Blue
spectra are allowed when the rate of variation of the external volume
is much smaller than the internal  one. The maximal $n$ achievable in
this case is $n=2$, corresponding to the case of static external manifold
($r \to \infty $).

Bearing in mind Eqs. (\ref{H1}) and (\ref{index}), we can 
use Eq. (\ref{norm}) and perform
the integral of Eq. (\ref{CL1}). For  $-3 < n < 3$ the integral appearing 
in Eq. (\ref{CL1}) can be done analytically \cite{dur2} and the result is 
\begin{equation}
C_\ell^{(SW)} =  \frac{2^{n}}{72} ~ f^2(\sigma_{\rm i}) \biggl(\frac{H_1}{M_{P}}\biggr)^2 
\biggl(\frac{\omega_0}{\omega_1}\biggr)^{n -1}  
 \frac{\Gamma(3-n)\Gamma(\ell-\frac{1}{2}+\frac{n}{2})}{
\Gamma^2(2-\frac{n}{2})\Gamma(\ell+\frac{5}{2}-\frac{n}{2})}.  
\label{CL2}
\end{equation}
Here $\omega_0\simeq 10^{-18}$ Hz and  $\omega_1(t_0)=H_1a_1/a_0$
are, respectively, the proper frequencies corresponding to the present
horizon  scale and to the present value of the cut-off scale $k_1$:
\begin{eqnarray}
&&\omega_{1}(t_0) = (H_1~H_{\rm eq})^{1/2} \biggl(\frac{\Gamma}{m}\biggr)^{1/6} \sigma_{\rm i}^{-2/3} 
z_{\rm eq}^{-1},~~~~~~~~~~~\sigma_{\rm i} <1,
\label{ga1}\\
&& \omega_{1}(t_0) = (H_1~H_{\rm eq})^{1/2} \biggl(\frac{\Gamma}{m}\biggr)^{1/6} \sigma_{\rm i}^{1/2}
Z_{\sigma}^{-1} z_{\rm eq}^{-1},~~~~~~\sigma_{\rm i} >1.
\label{ga2}
\end{eqnarray}
(we have rescaled $\om_1$ taking into account the kinematics of the
various cosmological phases from $t_1$ down to $t_0$).
The factor
$Z_\sigma= (a_{\rm osc}/a_\sigma)$ denotes the amplification of the
scale factor during the phase of axion-dominated, slow-roll inflation,
for the case $\sg_{\rm i}>1$.  
Notice  that  $\omega_1(t_0)$
depends on the  mass,  on the initial amplitude of the
axion background and on the axion decay rate.
If the axion decays at a typical scale fixed by 
Eq. (\ref{decb}), Eqs. (\ref{ga1}) and (\ref{ga2}) lead to 
\begin{eqnarray}
\omega_1(t_0) &\simeq & 10^{29} \om_0 
\left(H_1\over  M_{\rm P}\right)^{1/2}
\left(m\over \sigma_{\rm i}^2 M_{\rm P}\right)^{
1/3},~~~~~~~~~~~~~~~~~~~\sigma_{\rm i} < 1,
\label{om1a}\\
&\simeq & 10^{29}\om_0
\left(\sigma_{\rm i} H_1\over  M_{\rm P}\right)^{
1/2} \left(m\over \ M_{\rm P}\right)^{
1/3}Z_\sigma^{-1},~~~~~~~~~~~~~~\sigma_{\rm i} > 1
\label{om2a}
\end{eqnarray}
(we have used $H_0 \simeq 10^{-6} H_{\rm eq} \simeq 10^{-60}M_{\rm P}$).
Hence, in spite of the fact that the initial axionic  spectrum does not have
any mass dependence, the mass appears again when computing the
amplitude of the spectrum at the present horizon scale $\omega_0$.

The amplitude of the Bardeen potential, on the other hand,  is
constrained by the COBE  normalization of the quadrupole coefficient
$C_2$, which in our case is given by  
\begin{equation}
C_2= 
\alpha^2_n f^2(\sigma_{\rm i}) 
\left(H_1\over M_{\rm P}\right)^2\left(\omega_0\over
\omega_1\right)^{n-1},  
\label{C2}
\end{equation}
where 
\begin{equation}
\alpha_n^2 = \frac{2^{n}}{72}{\Gamma(3-n) \Gamma\left({3+n\over
2}\right)  \over
\Gamma^2\left({4-n\over 2}\right) \Gamma\left({9-n\over 2}\right)}.
\label{alpha}
\end{equation}
Using the experimental result \cite{ban}
\begin{equation}
C_2=(1.9\pm 0.23)\times 10^{-10},
\label{expnorm}
\end{equation}
we are thus led to the bounds
\begin{eqnarray}
&&
\alpha^2_n f^2(\sg_{\rm i})\sigma_{\rm i}^{2(n-1)/3}
\left(H_1\over M_{\rm P}\right)^{(5-n)/2}
\left(m\over M_{\rm P}\right)^{-(n-1)/3} 10^{-
29(n-1)} \simeq 1.9 \times 10^{-10},
~~~~\sigma_{\rm i} < 1,
\label{C2a}\\
&&
\alpha^2_n f^2(\sg_{\rm i})Z_\sigma^{n-1} \sigma_{\rm i}^{(1-n)/2}
\left(H_1\over M_{\rm P}\right)^{(5-n)/2}
\left(m\over M_{\rm P}\right)^{-(n-1)/3} 10^{-
29(n-1)} \simeq 1.9 \times 10^{-10},\,~~\sigma_{\rm i} >1.
\label{C2b}
\end{eqnarray}
These constraints, imposed by the COBE normalization,
will be  discussed at the end of the 
present section, and combined 
with other theoretical constraints 
pertaining to the various models of background evolution. 

\subsection{Acoustic peak region}

In the previous discussion of the modes that are outside the horizon 
before decoupling, we have completely neglected the
possible scattering  of radiation with baryons. In fact, if we move to
smaller angular scales (i.e. typically to $\ell \gaq 100$),  the main  contribution
to the CMBR temperature  fluctuations comes from the oscillations of the various
plasma quantities, the so-called Sakharov oscillations \cite{sak}.
A correct approach to  this
problem is then to perturb consistently the Boltzmann equations for the
different  species of the plasma \cite{ks2,hs1,hs2}. Furthermore
it can be relevant 
to discuss the case of a smooth transition between radiation 
and matter dominated epochs. In such a context it
becomes difficult to provide an analytical description of the system and, 
in order to compute the patterns of the acoustic oscillations, 
we will indeed  present some numerical examples in the third part of the 
present section. 

It is however useful to emphasize that the phases of the Bardeen 
potential for the adiabatic mode of Eq. (\ref{bardint4}) determine 
not only the relative weight of the Sachs--Wolfe contributions, but also
the  specific phase of the oscillatory patterns at small scales in the 
temperature fluctuations. 
For scales $\ell \gaq 100$ the contribution to the temperature perturbations 
given in Eq. (\ref{SW}) is dominated by acoustic oscillations.
This aspect can be appreciated by looking at Eqs. (\ref{delrel})--(\ref{uma}) 
 in the limit $k\eta >1$, where the peculiar velocity 
of baryonic matter does not oscillate. Instead, from Eq. (\ref{delrel}),
we find that the terms $\delta_{\rm r}/4$ and $
 \Phi$, appearing in Eq. (\ref{SW}),
combine to give a single term oscillating like a cosine:
\begin{equation}
\frac{\Delta T}{T}(k,\eta_{0},\eta_{\rm dec}) \simeq
  \frac{1}{4}\delta_{\rm r}(k,\eta_{\rm dec}) + \Phi_{0}(k) \sim 
\frac{\Phi_{0}(k)}{3} \cos{\omega\eta_{\rm dec}}.
\end{equation}
In this argument the interactions of baryons 
with the radiation fluid have been neglected.
The dynamics of $(\Delta T/T)_{k}$ can be obtained from an exact 
Boltzmann  equation with source term provided by Compton scattering coupled 
to the  continuity 
and Euler equations for the fluid variables. 
Before recombination, 
Compton scattering is very rapid and therefore  
the Boltzmann, Euler and continuity equations for the 
photon--baryon system can be expanded in powers of the Compton 
scattering  time \cite{hs1,hs2}. 
Within this approximation the baryon velocity 
field is damped and $(\Delta T/T)_{k}$ oscillates 
as a cosine for adiabatic initial conditions.
In the approximation 
of \cite{hs1,hs2}, the  oscillations
in $(\Delta T/T)_{k}$ have an amplitude 
 proportional to  $(1 + R)^{-1/4}$
where $R(\eta) = 3 \rho_{\rm b}/( 4 \rho_{\rm r})$. This 
result simply tells that the baryonic content 
of the plasma determines the height of the first peak.
Notice that this is in sharp contrast with what happens in the 
case of light axions \cite{dgs1,dgs2}, where the Bardeen potential is
quadratic in the axion fluctuations, and the initial
conditions for the hydrodynamical evolution are of the isocurvature type.
This implies, in particular,  that  the oscillatory patterns of the CMBR
anisotropies will be shifted by $\pi/2$  if compared with the case
discussed in the present paper.

\subsection{Constraints on pre-big bang models}

In this subsection we will discuss the bounds imposed by the COBE
normalization, together with other constraints following from the
evolution of the background geometry. Let us start with the axion
spectrum of minimal pre-big bang models, Eq. (\ref{norm}). In such a 
case, and for a flat Harrison--Zeldovich spectrum (i.e.  
$n=1$), the COBE normalization is inconsistent with a cut-off $H_1$ at
the standard value $M_{\rm s} \sim 10^{-1} ~M_{\rm P}$ of the string mass
scale \cite{kap}. By using $n=1$, and taking for $\sg_{\rm i}$ the value
minimizing $f(\sg_{\rm i})$, 
\begin{equation}
\sigma_{\rm i}^{\rm min} = \sqrt{\frac{c_2}{c_1}}\simeq 1.38,~~~~~~~~
f(\sigma_{\rm i}^{\min}) \simeq 0.34,
\end{equation}
we have indeed, from Eqs. (\ref{C2})--(\ref{expnorm}), 
\begin{equation}
H_1 \simeq 5.2 \times 10^{-4}~M_{\rm P}.
\end{equation} 
However, the precise value of $H_1$ is one of the main uncertainties 
of pre-big bang models. As we shall see in a
moment, the value $H_1 = M_{\rm s}$ (or $H_1 = M_{\rm GUT}$) may become consistent with the
COBE normalization for non-flat (blue) spectra, and even for a strictly
flat spectrum in the case of non-minimal implementations of the pre-big
bang scenarios.
 
Let us first recall the various constraints to be imposed on the
spectrum. The condition (\ref{C2a}) is to be combined with the constraint
(\ref{c1}),  the condition (\ref{C2b}) with the constraint (\ref{c2}), which
are required for the consistency of the corresponding classes of
background evolution. Both conditions are to be intersected with
the experimentally allowed range of the spectral index. We will use (as
a reference value) the generous upper bound \cite{cobe}, $ n \laq
1.4$. Also, for our illustrative purpose, we will take the
maximally extended range of allowed values of the axion
mass, satisfying the nucleosynthesis constraint $m \gaq 10$
TeV.

We will assume, finally, that in the case $\sg_{\rm i} >1$ the axion-driven
inflation is short enough, to avoid a possible contribution to $C_{\ell}$ 
arising from the metric fluctuations directly amplified from the
vacuum during such a phase of axionic inflation. This requires that the
smallest amplified frequency mode $\om_\sg$, crossing the horizon at
the beginning of inflation, at decoupling be still larger than the Hubble 
horizon  at the corresponding epoch. This imposes the condition 
 $\om_\sg(t_0) =H_\sg
(a_\sg/a_0)>\om_{\rm dec}(t_0)=H_{\rm dec}(a_{\rm dec}/a_0)$ ,
namely  
\begin{equation} Z_\sg \laq 10^{27} \sg_{\rm i} \left(m\over
M_{\rm P}\right)^{5/6}, 
\label{648}
\end{equation}
to be added to the constraint (\ref{c2}) for $\sg_{\rm i} >1$. 

\begin{figure}
\centerline{\epsfxsize = 11 cm  \epsffile{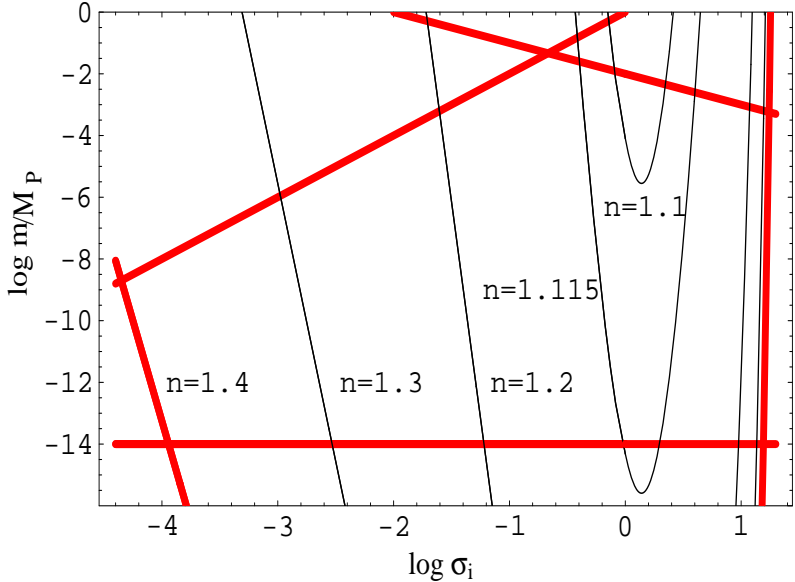}}
\vskip 3mm
\caption{ Allowed values of $\sg_{\rm i}$ and $m$ (in Planck units)
according to Eqs. (\ref{C2a}), (\ref{C2b}), with $H_1=10^{-2}M_{\rm P}$. 
The allowed region (within the thick lines) is bounded by the condition
$n<1.4$ (left and right bold lines), by the nuclesosynthesis constraint
$m>10$ TeV (lower bold line), by the condition (\ref{c1}) (upper left bold
line) and (\ref{c2}) (upper right bold line).}
\label{f9}
\end{figure}

The allowed region in the plane $\{\log \sg_{\rm i}, \log (m/M_{\rm P})\}$ is
illustrated in Fig. \ref{f9} for $H_1=10^{-2}M_{\rm P}$, 
using for the inflation
factor the  parametrization $Z_\sg =
\exp((\sg_{\rm i}^2-1)/8)$.  Along the thin full curves the parameters satisfy
the COBE normalization, for fixed values of $n$, ranging from $1.1$ to
$1.4$ (the condition (\ref{648}), in this case, is always automatically
satisfied). A growing (``blue") spectrum is  thus allowed even if
$H_1\sim M_{\rm s}$, for a wide range of axion masses, and for a (narrower)
range of values of  $\sg_{\rm i}$. In particular, for the case $H_1=10^{-2}
M_{\rm P}$, we find  $1 \gaq \sg_{\rm i} \gaq 10^{-4}$, for $\sg_{\rm i}<1$. For 
$\sg_{\rm i} >1$ the results are complementary for the spectral index, but
there are much more stringent bounds on $\sg_{\rm i}$, because the
inflationary red-shift factor $Z_\sg$ grows exponentially with
$\sg_{\rm i}^2$. As a consequence, the allowed region for
$\sg_{\rm i}>1$ is distorted and compressed, as illustrated in Fig. \ref{f9}. 

\begin{figure}
\centerline{\epsfxsize = 12 cm  \epsffile{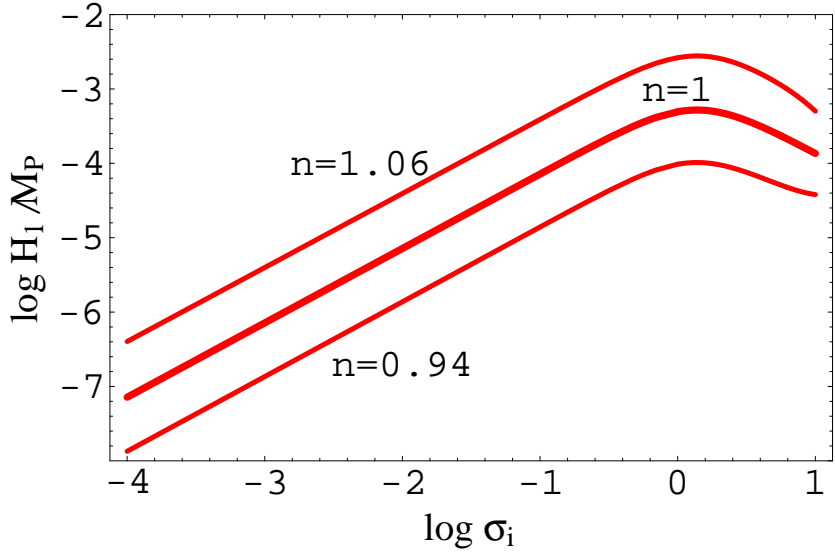}}
\vskip 3mm
\caption{Allowed values of $H_1$ as a function of $\sigma_{\rm i}$ 
for different values of the spectral index and for $m = 10^{-9} M_{\rm P}$. }
\label{f9a}
\end{figure}

The allowed region may be further extended if the inflation scale $H_1$
is lowered (see for instance \cite{sc}), and a flat ($n=1$) or
almost flat spectrum may become possible if $c_2\a_1 H_1 \laq
10^{-5}M_{\rm P} \sg_{\rm i}$, for $\sg_{\rm i}<1$, and if $c_1\a_1H_1 \laq
10^{-5}M_{\rm P} /\sg_{\rm i}$, for $\sg_{\rm i}>1$ (see Eqs. (\ref{C2a}),
(\ref{C2b})). The corresponding allowed values of $H_1$ and $\sg_{\rm i}$ are
illustrated in Fig. \ref{f9a} for $m=10^{-9}M_{\rm P}$, and for three
different values of $n$ around $1$.

However, a flat spectrum
may be allowed even keeping pre-big bang inflation at the string scale
($H_1 \sim M_{\rm s}$), provided we consider a non-minimal pre-big bang
scenario. In that context, in fact, the high-frequency branch of the
axionic spectrum may be modified, getting steeper enough to match the
string-scale normalization at the end-point of the spectrum, while 
the low-frequency branch  remains flat (or 
quasi-flat, see Appendix), to agree with large-scale observations. Examples of
realistic pre-big bang backgrounds producing such an axion spectrum
have been presented already in \cite{dgs3}. 

A non-minimal spectrum can be parametrized by  the Bogoliubov
coefficients (which will be given in Eq. (\ref{delta})), in terms of a generic 
break-scale $k_s$ and of the high-frequency slope parameter $\delta$. 
In that case, for a  long and/or steep enough high-frequency branch
of the spectrum, the large-scale amplitude may be suppressed
sufficiently to allow flat (or even red)  spectra at the COBE scale. 
In fact, for the non-minimal spectrum (\ref{delta}), the normalization
condition  (\ref{C2}) becomes 
\begin{equation}
C_2=
\alpha^2_n f^2(\sigma_{\rm i})
\left(H_1\over M_{\rm P}\right)^2\left(\omega_0\over
\omega_1\right)^{n-1}\left(\om_s\over \om_1\right)^\da.
\label{649}
\end{equation}
Flat or red spectra ($n \leq 1$) are thus possible even for $H_1 \gaq
10^{-2}M_{\rm P}$, provided
\begin{equation}
\alpha^2_1 f^2(\sigma_{\rm i})
\left(\om_s\over \om_1\right)^\da \laq 10^{-6}.
\label{649a}
\end{equation}

\begin{figure}
\centerline{\epsfxsize = 12 cm  \epsffile{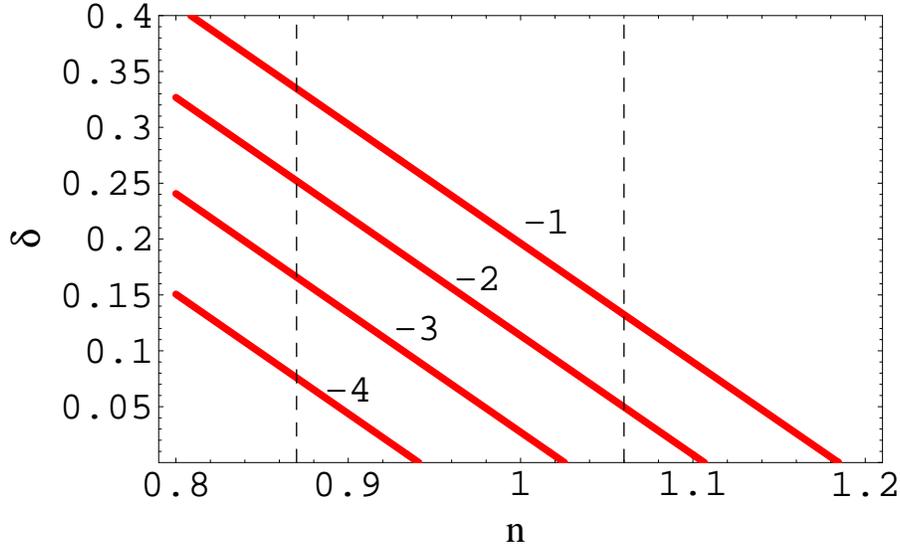}}
\vskip 3mm
\caption[a]{Relation between $\da$ and $n$ for different values of
$H_1$ (in Planck units), for $m=10^{-9}M_{\rm P}$, and for $\sg_{\rm i}=1$. 
The vertical dashed lines denote the experimentally allowed range
$0.87 \leq n\leq 1.06$.}
\label{f10}
\end{figure}

\begin{figure}
\centerline{\epsfxsize = 12 cm  \epsffile{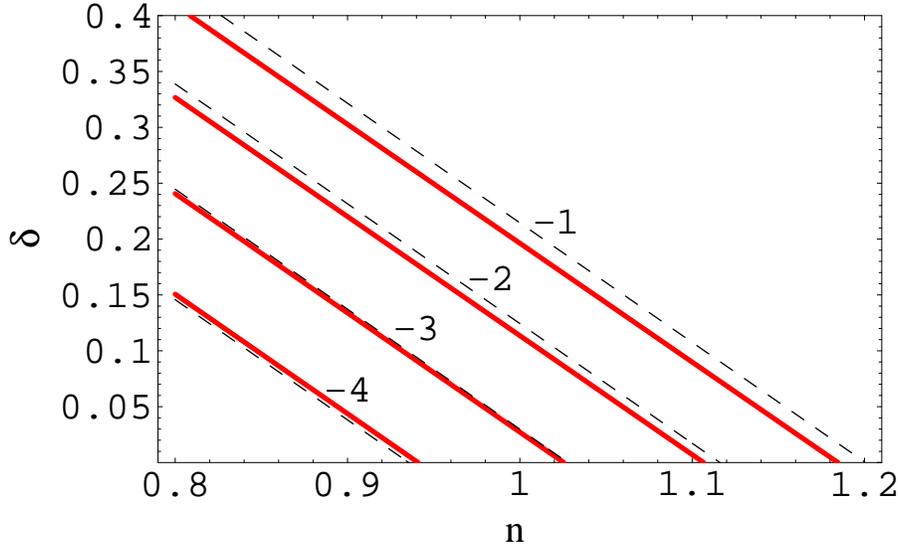}}
\vskip 3mm
\caption[a]{Stability of the curves of Fig. 10 for two different choices
of the axion mass, $10^{-9}M_{\rm P}$ (bold curves) and 
$10^{-14}M_{\rm P}$ 
(thin dashed curves).} 
\label{f11}
\end{figure}

In order to illustrate this possibility we will choose a specific model of
background by identifying $k_s$ with the equilibrium scale $k_{\rm
eq}$, in such a way that $n$ corresponds to the spectral index of all
scales relevant to the CMBR anisotropies, while $n+\da$ provides the
average spectral index for all other scales, up to $k_1$. We will also
assume for the axion background the ``natural" initial value $\sg_{\rm i}=1$,
so that 
\begin{equation}
\frac{\om_1}{\om_s}=\frac{\om_1}{\om_{\rm eq}}\simeq 10^{27}
\biggl(\frac{H_1}{M_{\rm P}}\biggr)^{1/2} 
\biggl(\frac{m}{M_{\rm P}}\biggr)^{1/3}.
\label{650}
\end{equation}
The COBE normalization can then be written explicitly as
\begin{equation}
C_2 = \alpha_{n}^2 f^2(1)
\biggl(\frac{H_1}{M_{P}}\biggr)^{(5-n - \delta)/2} 
\biggl(\frac{m}{M_{P}}\biggr)^{- ( n -1 + \delta)/3} 10^{-[27\delta + 
29(n-1)]}.
\label{C2break}
\end{equation}
By using the experimental value of $C_2$ given in Eq. (\ref{expnorm})
we can now obtain a relation between the high-frequency
slope parameter $\da$ and the spectral index $n$ at the COBE scale, for
any given value of $H_1$ and $m$. In Fig. \ref{f10} we illustrate such a
relation for different (realistic) values of $H_1$, and for a typical axion
mass $m=10^{-9}M_{\rm P}$. It should be stressed that, for $n \simeq
1$, and $\sg_{\rm i}$ of order 1 (i.e. near the minimum of $f(\sg_{\rm i})$),
the curves at constant $H_1$ are almost insensitive to the values of
$m$, and remain stable even if we change $m$ by various orders of
magnitude, as illustrated in Fig. \ref{f11}.

We have also reported, in Fig. \ref{f10}, the (present) most stringent 
bounds on $n$, obtained by a recent analysis of the CMBR anisotropies
and  large-scale structures \cite{ber,wtz}, i.e. $0.87 \leq n\leq 1.06$.
They are all compatible with $H_1 \simeq M_{\rm s}$, provided we allow for a
small break of the minimal spectrum, with $\da \simeq 0.2$--$ 0.3$. On
the other hand, as already stressed, no break at all is needed (i.e. $\da
=0$) if, for some dynamical mechanism (see for instance \cite{sc}) the
string mass is lowered down to the GUT scale, i.e. $H_1 \simeq
10^{-3}M_{\rm P}$. 

In Fig. \ref{f11} we have plotted the same curves of Fig. \ref{f10}  for
two, very different values of the axion mass, $10^{-9}M_{\rm P}$ (bold
curves) and $10^{-14}M_{\rm P}$ (thin dashed curves). As clearly
illustrated by the figure, the dependence on the mass is very mild, and
it becomes practically inappreciable (for the given range of parameters)
when $H_1$ approaches $M_{\rm GUT}$. 

\begin{figure}[ht]
\centerline{\epsfxsize = 12 cm  \epsffile{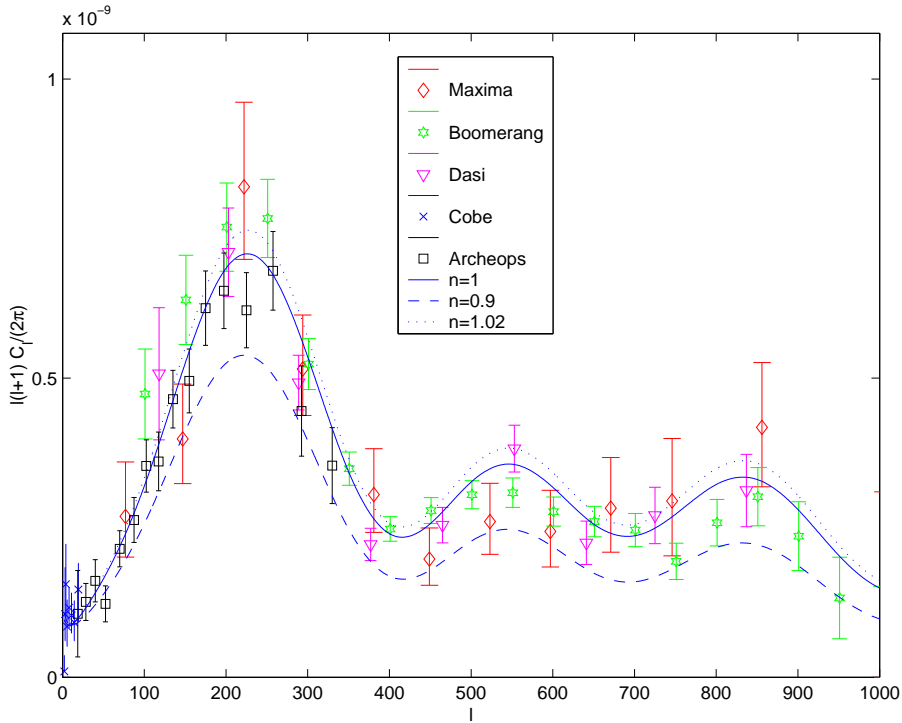}}
\vskip 3mm
\caption[a]{The  spectrum of $C_{\ell}$ is illustrated 
for a fiducial set of parameters ($h_0 =0.65$, $\Omega_{\rm b}=
0.04733$,  $\Omega_{\Lambda} = 0.7$, $\Omega_{\rm m} = 0.25267$) 
and for  flat (full line, $n =1$), slightly red (dashed line, $n =0.9$)
 and slightly blue (dotted line, $n =1.02$) spectral indices.}
\label{SR12a}
\end{figure}

\begin{figure}
\centerline{\epsfxsize = 12 cm  \epsffile{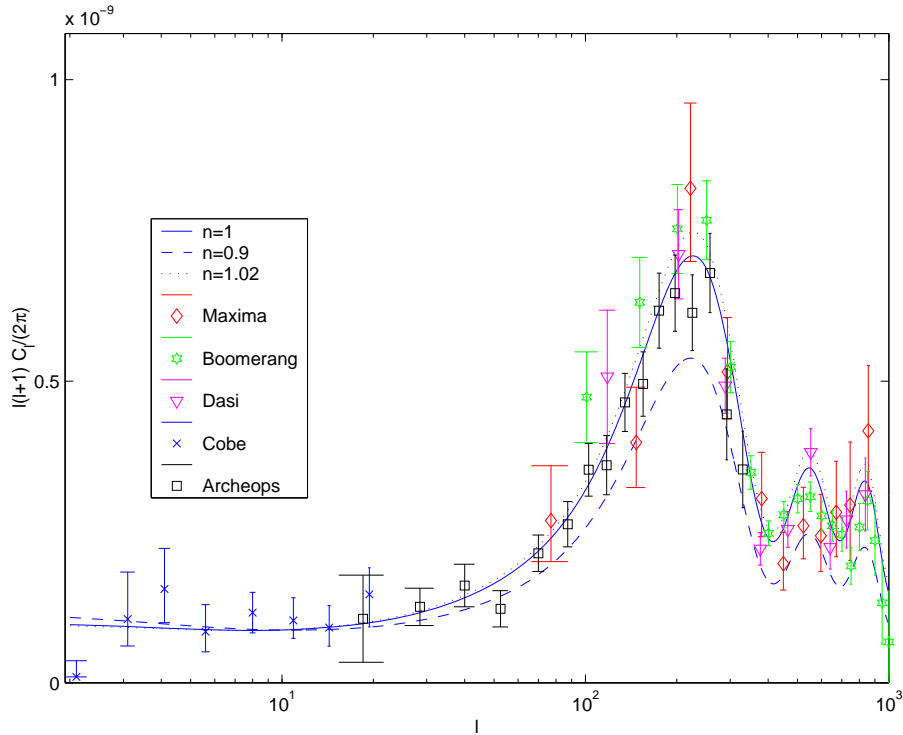}}
\vskip 3mm
\caption[a]{The same plot as in Fig. \ref{SR12a}, but 
on a semi-logarithmic scale.}
\label{SR13a}
\end{figure}

Having discussed the constraints imposed by the 
COBE normalization we can present now the plots of the angular
coefficients $C_{\ell}$ for the scalar spectrum, in the case of a spatially
flat background and  for few relevant choices of
the spectral index. In Figs. \ref{SR12a} and \ref{SR13a} 
the (scalar) angular power spectrum defined in Eq. (\ref{anps}) is reported
 for a flat, slightly red and slightly blue spectrum.
In order to obtain the results of  Figs. \ref{SR12a} and \ref{SR13a}
we used the latest release of CMBFAST, relaxing the strict 
COBE normalization at $\ell =10$, in favour of a better general agreement of the 
overall fit.

In Figs. \ref{SR12a} and \ref{SR13a} we used  the following
values of the cosmological parameters, selected according to  fits of 
CMBR  anisotropy experiments \cite{wtz}:  $h_{0} = 0.65$,
$\Omega_{\Lambda}= 0.7$, and $ h_0^2\Omega_{b} = 0.02$. The selected
value  of $h_{0}^2\Omega_{b}$ is rather robust, even if no final
consensus  has been reached on the second significant figure beyond
$0.02$.  We have also assumed the simplest scenario for the
late-time cosmological evolution, with no significant effects of 
reionization.

In Fig. \ref{SR12a} the  $C_{\ell}$ are plotted on a linear
scale, whereas in  Fig. \ref{SR13a} we present the same plot with 
a semi-logarithmic  scale, in such a way that  the region relevant to
the COBE observations is  less compressed.  
We recall, finally, that the flat, red and blue
spectral indices may correspond to particular combinations of the
parameters $H_1,m,\da$ and $\sg_{\rm i}$, chosen in such a way as to satisfy
the COBE normalization, Eq. (\ref{649}).  

The data points reported in Figs. \ref{SR12a} and \ref{SR13a} 
are those from COBE \cite{cobe,cob2}, BOOMERANG \cite{boom},
DASI \cite{dasi}, MAXIMA \cite{maxima} and ARCHEOPS \cite{archeops}.
Notice that the data reported in \cite{archeops} fill 
the ``gap'' between the last COBE points and the points of 
\cite{boom,dasi,maxima}. Therefore, one could think of normalizing the 
spectra  not to COBE but directly to ARCHEOPS. 
In spite of this possibility, the forthcoming MAP data 
will give even more accurate determination of the $C_{\ell}$ spectra. It 
will then be interesting to use these data in order to make 
more consistent and accurate determinations of the pre-big bang 
parameter  space.

\renewcommand{\theequation}{7.\arabic{equation}}
\setcounter{equation}{0}
\section{Concluding remarks}

In the present paper the possible conversion of isocurvature,
primordial axionic fluctuations into adiabatic, large-scale metric
perturbations has been discussed in the context of the pre-big bang
scenario.  Depending upon the specific relaxation of the axionic 
background toward the minimum of the potential, a constant (and
large enough) mode  in the Bardeen potential can be generated, for
scales  that are still outside the horizon right after matter-radiation 
equality. 

After analysing the dynamics of the background and of its fluctuations,
the final  amplitude and spectrum of the Bardeen potential 
has been related to the initial axion spectrum 
directly arising from the vacuum fluctuations amplified during the  
pre-big bang epoch. Our goal has been to include, with reasonable 
accuracy, the details of the post-big bang evolution, in  such a way that 
the pre-big bang parameters could be directly constrained 
by the COBE normalization, and by the analysis of the Doppler-peak
structure.  All the theoretical uncertainty reflects in our lack 
of knowledge of $H_1$ which determines the end point of the primordal
axion spectrum.

The main conclusion of this work is that a phenomenologically appealing
spectrum of 
adiabatic scalar perturbations can naturally emerge from the simplest pre-big
bang
scenario through a conversion of the initial isocurvature perturbations of the
Kalb-Ramond axion.
Since, at the large scales tested by CMBR experiments, the above conversion
preserves 
the scale-dependence of the original spectrum, it is important for the latter
to be
quasi-scale-invariant at large scales. This can be achieved, for instance, if
  the very
early stages of pre-big bang cosmology at weak coupling involve a symmetric
evolution
of all $9$ spatial dimensions (modulo $T$-duality).
Since the constant mode of the curvature fluctuations leads to
adiabatic initial conditions for the fluid evolution after matter-radiation
equality, the location of the Doppler peak is correctly reproduced.

On the other hand, the absolute normalization of fluctuations at large scales
(say those
relevant for COBE) 
depend on several details of the model. Indeed, the axion spectrum is naturally 
normalized at its end-point, given by our parameter $H_1$.  
If one takes, naively, $H_1 \sim M_{\rm s} \sim 10^{17} ~{\rm GeV}$ and assumes a flat
spectrum
one finds values of $\Delta T/T$ that are a couple of orders of magnitude too
large
when compared with COBE's data.
However, one can think of many (individual or combined) effects that
 can bring down our normalization to agree with the data,
e.g.
\begin{itemize}
\item A slight (blue) tilt to the spectrum;
\item A blue spectrum just at high frequency 
(i.e. for scales that exit late, during the strongly coupled regime);
\item A lower  $H_1/ M_{\rm s}$  ratio;
\item A  lower $M_{\rm s}/ M_{\rm P}$ ratio.
\end{itemize}

In the near future we hope to extend the present discussion to 
forthcoming CMBR anisotropy data at even smaller angular scales. It would be interesting 
to see if a combined analysis of the experimental data 
may give further useful hints on the parameter space of the scenario 
explored in the present investigation. 

\section*{Acknowledgements}
V. Bozza would like to thank the ``Museo Storico della
Fisica e Centro Studi e Ricerche E. Fermi'' for financial support.
G. Veneziano wishes to acknowledge the support of a ``Chaire Internationale
Blaise Pascal'', administred by the ``Fondation de l'Ecole Normale
Sup\'erieure''. M. Giovannini is indebted to the 
``Institut de Physique Th\'eorique''
of the ``Universit\'e de Lausanne'' for partial support.

\newpage 

\begin{appendix}

\renewcommand{\theequation}{A.\arabic{equation}}
\setcounter{equation}{0}
\section{Axionic spectra}
During the pre-big bang phase the  quantum mechanical 
fluctuations of the axionic field will be amplified from the initial
vacuum state.  The obtained spectrum provides the initial
condition for the  evolution of the axion fluctuations in the post-big
bang phase.  At very large-scales, such a spectrum  will not depend so
much upon the details of the pre-big bang evolution.  At smaller scales,
however, it can be strongly affected by specific dynamics of the strong 
coupling and high-curvature regime.  In
spite of the fact that the spectral  slope at large scales is not  affected
by high energy corrections, the large scale amplitude is affected and,
in particular, a steeper slope at small scales has important
consequences for the 
normalization of the low-frequency branch of the spectrum.  In this
appendix we will consider, separately, the axion spectrum obtained
in the case of minimal  and non-minimal pre-big bang models.

\subsection{Minimal pre-big bang models}

The linearized evolution of massive axion 
inhomogeneities $\chi_k$, neglecting their coupling to scalar  metric
perturbations, in a spatially flat cosmological
background, is described in general by the equation 
\begin{equation}
\psi_{k}'' + \biggl[ k^2 + m^2 a^2 - \frac{z''}{z}\biggr] \psi_{k} =0,
\label{flucchi}
\end{equation}
where 
\begin{equation}
z = a~e^{\varphi/2}, ~~~~\psi_{k} = z\chi_{k}.
\label{pump}
\end{equation}
In the pre-big bang phase ($\eta < \eta_1$)  
the axion is massless. In the post-big bang, 
radiation-dominated phase, taking place for $\eta> \eta_1$,
 the gauge coupling freezes ($\varphi=$ const) and 
the axion acquires a mass. The produced 
axion spectrum, in principle, has a relativistic  and a 
non-relativistic branch: this is because, in the radiation 
era, the proper momentum is red-shifted
with respect to the rest mass, and the whole spectrum, mode by mode, 
tends to become non-relativistic. The spectral slope of the relativistic
and non-relativistic branches  of the spectrum are in general different.
However, if the  axion modes, as in the present case,  become 
non-relativistic when they are still outside the horizon, 
the  solution is then
exactly the same as in the relativistic limit.

Consider first the relativistic branch of the spectrum.
For $\eta <\eta_1$ the solution of Eq. (\ref{flucchi}) can be expressed 
 in terms of the
second-kind Hankel functions \cite{abr} as:
\begin{equation}
\psi_k(\eta)=\eta^{1/2}H_\mu^{(2)} (k\eta),
\label{psirel}
\end{equation}
 where $\mu$ depends on the parameters controlling the kinematics of
the pre-big bang background  (a specific example 
will be given below, see Eqs. (\ref{ckappa}) and (\ref{index})).  In the
radiation era, $\eta>\eta_1$, one has $z''/z=0$, and  the evolution
equation of $\psi_{k}$ acquires a massive  correction: 
\begin{equation}
\psi_k'' +\left(k^2 + m^2 a^2\right)\psi_k=0,
\label{psinr}
\end{equation}
Assuming that the axion mass is negligible at the transition epoch
$\eta_1$,   the solution (\ref{psirel}) can be matched to the
plane-wave solution
\begin{equation}
\psi_k= {1\over \sqrt k}\left[c_+(k) e^{-ik\eta}+
c_-(k) e^{ik\eta}\right],
\end{equation}
and the final result for $\chi_{k}$ is 
\begin{equation}
\chi_{k}( \eta) = {c(k)\over a \sqrt k}\sin (k\eta),
\label{chires}
\end{equation}
where 
\begin{equation}
c(k) \simeq \biggl(\frac{k}{k_1}\biggr)^{\frac{n - 5}{2}},
\label{ckappa}
\end{equation}
with $n = 4 - 2|\mu|$ . Note that 
the expression of the Bogoliubov coefficient $c(k)$ 
and of the mean  number of produced axions,  
$\overline{n}_{k} = |c(k)|^2$,
contains different numerical factors of order $1$. At the same 
time the maximal amplified momentum $k_1$ can be defined in different 
ways, all equivalent up to numerical factors. 
In the present analysis we will define the maximal scale $k_1$ as the
energy scale where one axion is produced per unit volume of phase-space.

Consider now the non--relativistic spectrum in the 
case when the mode becomes non--relativistic 
while it is still {\em outside the horizon}.
  Defining
as $k_m$ the limiting comoving frequency of a mode that becomes
non-relativistic ($k_m=ma_m$) at the time it re-enters the horizon
($k_m=H_ma_m$), we find, in the radiation era \cite{dgs2,dgs3}, 
\begin{equation}
k_m= k_1 \left(m\over H_1\right)^{1/2}.
\label{c11}
\end{equation}
We are thus considering modes with  $k\ll k_m$.
In order to estimate the spectrum, in this limit,  
let us write Eq. (\ref{psinr}) in a form 
suitable for comparison with known results 
of parabolic cylinder equations:
\begin{equation}
{d^2\psi_k\over dx^2}+\left({x^2\over 4} -b\right)\psi_k=0,
~~~~~ x=\eta (2 \alpha)^{1/2}, ~~~~ -b= k^2/2\alpha , 
\label{parab}
\end{equation}
where
\begin{equation}
m^2a^2 =\alpha^2\eta^2, ~~~~~~~~~~~~~~~~
\alpha= m H_1a_1^2,
\end{equation}
and where $a \sim \eta$ has been assumed. The corresponding 
general solution can be written as 
\begin{equation}
	\psi=A y_1(b,x)+B y_2(b,x)~,
\label{seceq}
\end{equation}
where $y_1$ and $y_2$ are the even and odd parts of the parabolic
cylinder functions \cite{abr}. The normalization to Eq. (\ref{chires})
in the relativistic limit (i.e. $x \to 0$) gives $A=0$ and
\begin{equation}
\psi_k \simeq c(k)\left(k\over 2\alpha \right)^{1/2} y_2(b,x) .
\end{equation}
Outside the horizon, $k\eta \ll1$, and for non-relativistic modes, $k \ll
ma$, we take (respectively) the limits $-b x^2 \ll1$ and $-b \ll x^2$,
the solution can be expanded as $y_2 \sim x\sim \eta \sqrt{2\a}$, so
that the mass disappears from the amplitude:
\begin{equation}
|\chi_{k}| \simeq {|c(k)|k^{1/2} \over a_1\eta_1}.
\end{equation}
The insertion of the spectrum (\ref{ckappa}), using $k_1=a_1 H_1$,
leads to the final result
 \begin{equation}
k^{3/2} |\chi_{k}| \simeq H_{1}\biggl(\frac{k}{k_1} 
\biggr)^{\frac{n -1 }{2}}.
\label{estimate}
\end{equation}

\subsection{Non-minimal pre-big bang evolution
and spectral breaks}

Equation (\ref{estimate}) holds in the case of minimal 
pre-big bang models, where the dynamical evolution 
of the dilaton field is dictated by the  solution 
of the low-energy equations of motion. However, when the dilaton 
enters the strong coupling regime, different types of 
scenarios may emerge. In particular,  relation (\ref{pump})
defining the form of the axion pump field, may change in the infinite
bare  string-coupling limit, as suggested by the arguments recently
developed in \cite{sc}. In the framework of  \cite{sc}
the axion coupling function, as well as the other coupling functions 
pertaining  to fields of different spin, may have a finite limit for 
infinite  bare string coupling. Hence, toward the end 
of the pre-big bang phase (i.e. when strong coupling is presumably 
reached), \begin{equation}
z \sim a [c_{z} + {\cal O}(e^{- \varphi/2})],
\end{equation}
where $c_{z}$ is a constant. Since the axionic pump field now depends only on
 the  scale factor, it will  naturally be steeper for small length
scales.  A complementary possibility, discussed in 
\cite{dgs3}, is the presence of an intermediate high-energy phase,
which precedes the standard radiation era, and which is still part of the
accelerated pre-big bang regime, but in which the kinematics of the
(usual) canonical pump field is significantly different from its
low-energy behaviour.

In all these cases the obtained spectra, at small scales, are possibly 
steeper than in  the case of minimal pre-big bang models. 
In the simplest case the spectrum will have only one break,  
at a momentum scale that will be conventionally denoted by $k_{s}$,
and  the Bogoliubov coefficients can be written in the form
\begin{eqnarray}
|c_k|^2 &=& \left(k\over k_1\right)^{n- 5 + \delta},
~~~~~~~~~~~~~~~~~~~~~~k_s<k<k_1,
\nonumber\\
&=& \left(k_s\over k_1\right)^{n- 5 + \delta}
\left(k\over k_s\right)^{n-5},
~~~~~~~~~~~~ ~~k<k_s. 
\label{delta}
\end{eqnarray}
Here $\delta > 0 $ parametrizes the slope of the break at high
frequency, while $n$ is the usual spectral index appearing at 
large scales and computed on the basis of the perturbative 
evolution of the dilaton field. 
From Eq. (\ref{delta}) it can be argued that 
the steeper and/or the longer the
high-frequency branch of the spectrum, the larger the suppression
at low-frequency scales.

\end{appendix}

\end{document}